\documentclass{aa}  
\usepackage{array}
\usepackage{textcomp, gensymb}
\usepackage{comment}
\usepackage{xcolor}
\usepackage{footmisc}
\usepackage{tablefootnote}
\usepackage{float}
\usepackage{commath}
\usepackage{calrsfs}
\DeclareMathAlphabet{\pazocal}{OMS}{zplm}{m}{n}
\newcommand{\Ib}{\pazocal{I}}
\usepackage{graphicx}
\usepackage{txfonts}
\usepackage{hyperref}
\hypersetup{colorlinks,linkcolor={blue},citecolor={blue},urlcolor={blue}}
\begin{document} 
   \title{Parameter study for hot spot trajectories around Sgr$A*$}
   \author{Eleni Antonopoulou\inst{1}\fnmsep\inst{2} and Antonios Nathanail\inst{2}}
   \institute{Department of Physics, National and Kapodistrian University of Athens, University Campus, GR 15784 Zografos, Greece
    \and
    Research Center for Astronomy and Applied Mathematics, Academy of Athens, Soranou Efesiou 4, 115 27 Athens, Greece
             }
   \date{Received April 30, 2024; accepted July 2, 2024}
  \abstract
  % context heading (optional) 
   {Intense flaring events in the near-infrared and X-ray wavebands of our Galactic center have been the subject of research for decades. In recent years, the GRAVITY instrument of the Very Large Telescope captured the motion and polarimetric signature of such a flare in close proximity to the supermassive black hole.} 
  % aims heading (mandatory)
   {This study aims to investigate a broad parameter space for hot spot motion in the vicinity of Sgr$A*$ and reproduce the observed flaring behavior.}
  % methods heading (mandatory)
   {To this end, we have developed a general relativistic radiative transfer code and conducted a parameter study including both planar and ejected hot spot configurations around supermassive black holes.}
  % results heading (mandatory)
   {Super-Keplerian orbital frequencies are favored by circular equatorial, cylindrical and parabolic models, whereas conical hot spot trajectories provide a better fit for orbital frequencies below the Keplerian value. Additionally, a distant observer cannot effectively differentiate between Schwarzschild and Kerr black holes, as well as face-on orbits at different observation angles.}
  % conclusions heading (optional)
  {}
   \maketitle
%
%--------------------------------------------------------------
\section{Introduction}
Extended monitoring of stellar orbits in very close proximity to our Galactic center (\citealt{genzel2010}; \citealt{genzel2021}) has revealed the existence of a supermassive black hole with a mass of $M\simeq4.1\times 10^6\,M_{\odot}$ and distance of $D\simeq 8.178$ kpc (\citealt{Schodel2002}; \citealt{Ghez_2003}; \citealt{Gillissen2009}; \citealt{GRAVITY2019}). 
In its quiescent state, Sagittarius $A*$ exhibits a bolometric luminosity of $\sim 5\times 10^{35}$ erg s$^{-1}$, almost ten orders of magnitude below its Eddington luminosity, while its accretion rate reaches $\sim 10^{-8}$ M$_{\odot}$ yr$^{-1}$ (\citealt{Quataert_2000}; \citealt{Marrone_2007}; \citealt{EHT2022}), rendering it one of the faintest sources to date. \par   
The Galactic center also exhibits strong flaring events in the near-infrared (NIR) and X-ray wavebands several times a day. 
The NIR flares produce ten times more luminosity than the quiescent state (such flares can reach flux levels of $\sim 6$ mJy (\citealt{Do2019})) 
and are strongly polarized with a changing polarization angle (\citealt{genzel2003, Eckart2006, trippe2007}).  
One in four NIR flares is also followed by an X-ray counterpart that can reach an increase in luminosity of up to two orders of magnitude (\citealt{baganoff2001, porquet2003, hornstein2007}). 
While the high linear polarization degree of NIR flares (up to $\sim40\%$) implies a synchrotron source of emission, the mechanism governing X-ray flaring events is still widely unknown (\citealt{Yuan_2003, ponti2017}). \par
In recent years, the GRAVITY Collaboration has detected several NIR flares in the vicinity of Sgr$A*$, providing both astrometric and polarimetric observations with the GRAVITY interferometer at the Very Large Telescope Interferometer (\citealt{GRAVITY_2018}, \citealt{GRAVITY2023}).
In all cases, the observed flares exhibit a clockwise motion over a period of approximately one hour, during which the polarization vector completes one full loop (\citealt{GRAVITY2023}). 
The GRAVITY Collaboration calculated the flux-centroid positions of the corresponding hot spots with an astrometric accuracy of $\sim2\,r_g$ (\citealt{GRAVITY2017}) and employed circular Keplerian models to fit the well-resolved flares across various orbital radii and observer inclinations (\citealt{GRAVITY_2018}; \citealt{GRAVITY2020}). 
In their latest work, the GRAVITY Collaboration also investigated the light curves and polarized signatures of these hot spot trajectories, and set boundaries on both orbital shear and nonplanar motion (\citealt{GRAVITY2020}; \citealt{GRAVITY2020polarimetry}). \par
Although investigating the motion and flare emission of hot spots orbiting the supermassive black hole in our Galactic center has been the subject of research for more than a decade (\citealt{Broderick2005}; \citealt{Ziri2015}), the data published in \citealt{GRAVITY_2018} shed new light on the topic. Numerous analytical and semi-analytical plasmoid models have since been proposed to interpret both the astrometric flux-centroid positions (\citealt{Ball_2021}; \citealt{Matsumoto_2020}; \citealt{Lin2023}; \citealt{Aimar2023}; \citealt{Huang2024}; \citealt{kocherlakota2024}) and the polarized signatures (\citealt{Vos2022}; \citealt{vincent2023}; \citealt{yfantis2023}) of the NIR flares. On the numerical regime, state-of-the-art general relativistic magnetohydrodynamic (GRMHD) simulations have predicted and investigated hot spot formation in highly magnetized accretion disks surrounding supermassive black holes (\citealt{Dexter2020, Chatterjee2021, Porth2021, Nathanail2022}). Flux eruption events in magnetically arrested disks are expected to produce flaring activity (\citealt{Ripperda2022}), however, this may not fit the GRAVITY results well due to their sub-Keplerian motion (\citealt{Porth2021}). The flaring events of 2018 have also motivated research into alternative theories of gravity and the nature of the compact object in our Galactic center (\citealt{Shahzadi_2022}; \citealt{Rosa_2022}; \citealt{Chen2024}). \par
Despite the recent surge in black hole research, the motion of the observed flares has not been effectively reproduced to this day. 
Although the circular Keplerian orbits presented in \citealt{GRAVITY_2018} and \citealt{GRAVITY2020} provide a promising initial fit, their flux-centroid positions consistently fall within the interior of the observed error bars, in order to account for the desired orbital period. 
When modeling the July 22, 2018 flare, \citealt{Matsumoto_2020} show that considering larger orbital frequencies and moving the orbital plane yields significantly better results while sacrificing the value of the goodness of fit parameter. 
Moreover, both the coronal mass ejection (\citealt{Lin2023}) and magnetic reconnection models (\citealt{Aimar2023}) demonstrate specific hot spot trajectories that are in good agreement with the astrometric positions and light curve of the July 22 flare.
\par
This work aims to explore the parametric space of plasmoid orbits in the vicinity of a supermassive black hole and fit the observations of \citealt{GRAVITY_2018} for Sgr$A*$. 
In more detail, we explore circular, cylindrical, conical and parabolic trajectories for varying orbital frequency and observer inclination. 
Furthermore, we study the impact of the black hole spin on these families of orbits.
To this end, we have developed a  general relativistic radiative transfer (GRRT) Python code to study the kinematics of hot spots orbiting Sgr$A*$.  
We provide an overview of our radiative transfer scheme in Section \ref{Sect:Num.Setup}, while several tests that were incorporated to assess the validity of our results are presented in Appendix \ref{Append:Code_Eval}.\par
The rest of the paper is organized as follows.  
In Section \ref{Sect:Circular}, we study circular hot spot trajectories and investigate the effect of the black hole's spin and observation angle on the reconstructed image. 
In Section \ref{Sect:Cylind} a cylindrical hot spot model is adopted, whereas in Section \ref{Sect:Conical} conical trajectories are employed and in Section \ref{Sect:Parab} we study parabolic orbits.
We present our conclusions in Section \ref{Sect:Concl}.
\section{Numerical setup}\label{Sect:Num.Setup}
We first defined the observer's position with respect to the black hole, which is located at the center of our coordinate system. The metric for a Kerr black hole in Boyer-Lindquist coordinates is given by
\begin{align}
    \label{kerr_metric}
    ds^2=-&\left(1-\frac{2Mr}{\Sigma} \right)\,dt^2-\frac{4Mar\sin^2\theta}{\Sigma}\,dt\,d\phi
    +\frac{A}{\Sigma}\sin^2\theta\, d\phi^2 \nonumber\\ 
    +&\frac{\Sigma}{\Delta}dr^2+\Sigma\, d\theta^2 \,,
\end{align}
where $\Delta(r) \equiv r^2 -2Mr + a^2 \,,\: \Sigma(r,\theta) \equiv r^2 + a^2\cos^2\theta$ and \linebreak $A(r,\theta) \equiv (r^2 + a^2)^2 - a^2\Delta\sin^2\theta$. Then, we initiated the ray-tracing scheme for $N_p\times N_p$ photons with initial conditions on the image plane of the observer. Following the formulation of \citealt{Younsi_PhD_2014}\footnote{See equations (3.58)-(3.60) in the respective paper.}, the ray's initial velocity was chosen so that each photon arrives perpendicularly to the image plane.\par
To determine the ray's trajectory as it approaches the Kerr black hole, one must integrate the following system of ordinary differential equations (\citealt{Fuerst_2004, Younsi2012, Younsi_PhD_2014}):
\begin{align}
    \dot{t}=\:&E+\dfrac{2Mr}{\Sigma\,\Delta} \left[ (r^2+a^2)E -aL_z \right] \,, \label{eq_tdot}\\
    \dot{p_r} =\:&\dfrac{\,1\,}{\Sigma\,\Delta} \left[ \kappa\,(1-r) \right.
    +2r\,(r^2+a^2)\,E^2 - \left. 2aEL_z \right] - \dfrac{2p_r^2\,(r-1)}{\Sigma} \,, \label{eq_prdot} \\ 
    \dot{r}=\:&\dfrac{\,\Delta\,}{\Sigma}\,p_r \,, \label{eq_rdot} \\
    \dot{p_{\theta}} = \:&\dfrac{\sin\theta\,\cos\theta}{\Sigma} \left[ \dfrac{L_z^2}{\sin^4\theta} - a^2E^2 \right] \,, \label{eq_pθdot} \\
    \dot{\theta} = \:&\dfrac{\,1\,}{\Sigma}\,p_{\theta} \,, \label{eq_θdot} \\
    \dot{\phi}=\:&\dfrac{2MarE+(\Sigma-2Mr)\,L_z\csc^2\theta}{\Sigma\,\Delta} \,, \label{eq_φ}
\end{align}
where $E$ and $L_z$ are the photon's energy and axial angular momentum, and $\,\kappa \equiv p_{\theta}^2 + L_z^2\csc^2\theta + a^2E^2\sin^2\theta$. Furthermore, to calculate the intensity of the ray, one must also integrate the radiative transfer equations shown below, along the photon's geodesic
\begin{align}
    \frac{d\,\Ib}{d\lambda} &= \gamma^{-1}\left(\frac{\,j_{0,v}\,}{v_0^3} \right)\,e^{-\tau_{v}} \,, \label{dI/dλ} \\[6pt]
    \frac{d\tau_{v}}{d\lambda} &= \gamma^{-1}\,\alpha_{0,v} \,, \label{dτ/dλ} 
\end{align}
where $v$ is the frequency of the radiation, $\Ib\equiv I_{v}/v^3$ is the Lorentz-invariant specific intensity, $\tau_{v}$ is the optical depth, $j_{v}$ and $\alpha_v$ are the emission and absorption coefficients, and $\gamma$ is the photon's relative energy shift between the point of emission (from a fluid component with $4-$velocity $u^{\alpha}$) and the point of detection,
\begin{equation}
    \label{γ_rad}
    \gamma = \frac{v}{\,v_0\,} = \frac{\;\;\;k_{\beta}u^{\,\beta}\big|_{\lambda_{obs}}}{\,k_{\alpha}u^{\,\alpha}\big|_{\lambda}\,} = \frac{-E}{\,-E\,u^t+L_z\,u^{\phi}+g_{rr}\,k^ru^r+g_{\theta\theta}\,k^{\theta}u^{\theta}\,} \,.
\end{equation}
In the following models, we consider a spherical hot spot with a constant radius of $0.5\,{\rm r_g}$\footnote{\,The gravitational radius of the black hole is given by $r_g=GM/c^2$.}, where ${\rm r_g}\approx 6 \times 10^{11} \,{\rm cm}$ for Sgr$A*$, and 
employ a power-law electron distribution. 
In particular, the synchrotron emissivity for the electrons is given by (\citealt{Pandya_2016})
\begin{align}
    \label{eq:non_thermal_emissivity}
    j_{v}=\:\dfrac{n_e\,e^2\,v_c}{c}\,&\dfrac{3^{p/2}\,(p-1)\,\sin\theta}{2\,(p+1)\,(\gamma_{min}^{1-p} - \gamma_{max}^{1-p})} \nonumber \\
    &\Gamma\left(\dfrac{3p-1}{12} \right)\,\Gamma\left(\dfrac{3p+19}{12} \right)\left(\dfrac{v}{v_c\,\sin\theta} \right)^{-(p-1)/2} \,,
\end{align}
where $n_e$ is the number density of nonthermal electrons, $e$ is the electron charge, $v_c$ is the electron cyclotron frequency, $p$ is the power-law index, $\theta$ is the angle between the magnetic field vector and the emitted photon wave vector, and $\gamma_{max}$, $\gamma_{min}$ are the upper and lower limits of the particle Lorentz factor, respectively.
On the other hand, synchrotron absorption in the NIR regime is expected to be low and can be neglected at a first-order approximation, therefore, the investigated models do not include an ambient supermassive black hole accretion flow. \par
\begin{figure*}
\centering
\includegraphics[width=0.298\linewidth]{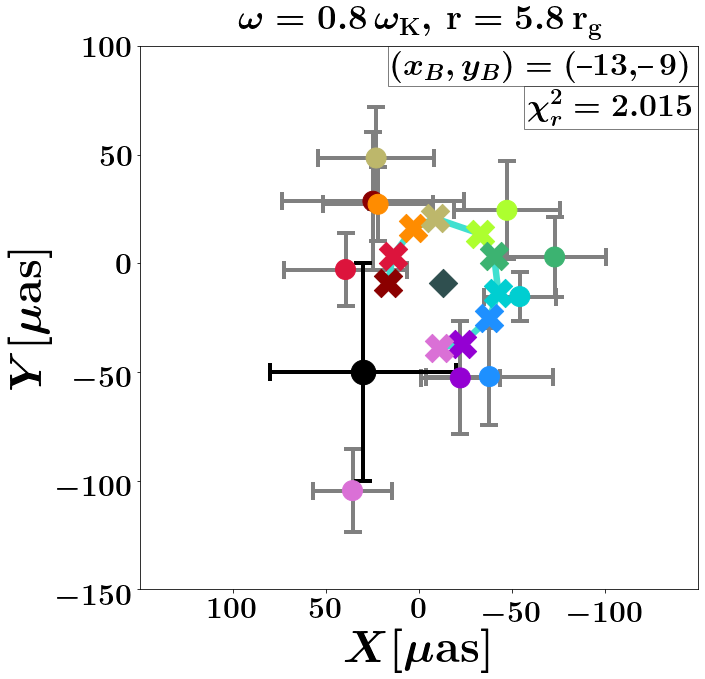} %
\hspace{0.24cm}
\includegraphics[width=0.278\linewidth]{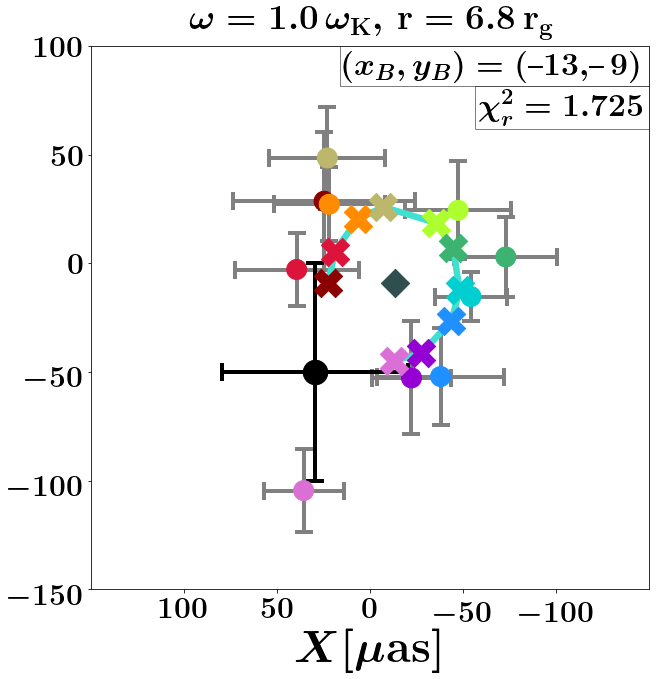}
\hspace{0.24cm}
%[width=0.3\linewidth]{GRAVITY_Plots/Circ_Kepler.png}
%\hspace{0.2cm}
\includegraphics[width=0.346\linewidth]{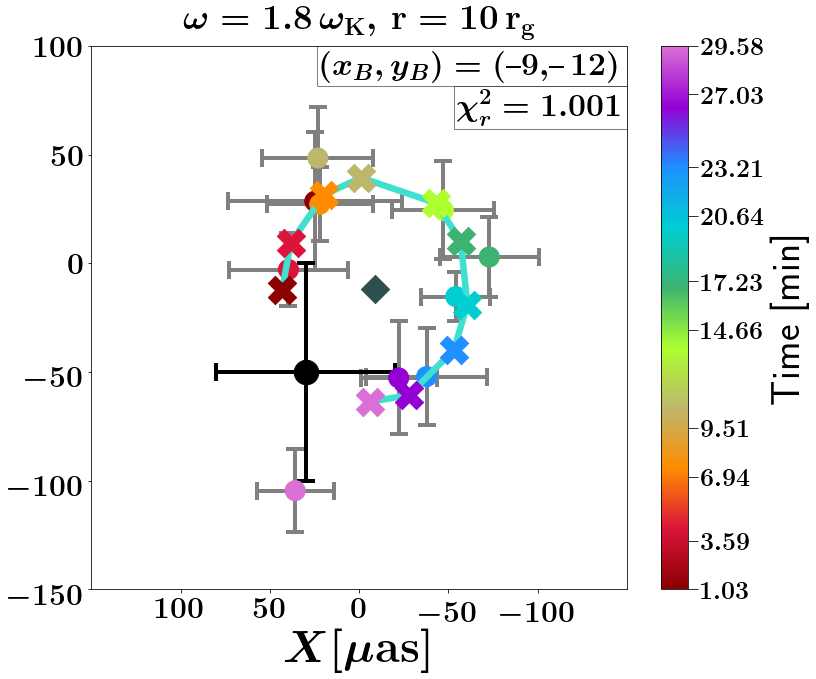}
\caption{Circular hot spot trajectories on the equatorial plane of a Schwarzschild black hole overlapped with the observed flare of July 22, 2018. Calculated flux-centroid positions (colorful 'x') are color-coordinated with the observations (colorful circles) and the associated timestamps are denoted in the far-right color bar. The approximate position of Sgr$A*$ in the sky, derived from observations, is depicted by a black circle. The black hole position for each orbit is illustrated by a dark gray diamond and denoted in the top right legend, along with the associated value of $\chi_r^2$. The observer angle is face-on and the orbital frequency of the hot spot corresponds to \textit{Left Panel}: Sub-Keplerian motion ($0.8\,\omega_K$), \textit{Middle Panel}: Keplerian motion ($\omega_K$), and \textit{Right Panel}: Super-Keplerian motion ($1.8\,\omega_K$), respectively.}
\label{Fig:Circular}
\end{figure*} 
Finally, we calculated the flux centroid position for each snapshot and evaluated the goodness of fit via the reduced $\chi^2$ parameter (\citealt{Matsumoto_2020}), whose formula is provided below:
\begin{equation}
    \label{χ^2}
    \chi_r^2 \,\equiv\, \dfrac{1}{\,2N-N_f\,} \sum_{i\,=\,1}^{N}\left[ \left(\dfrac{X(t_i)-X_i}{\sigma_{X_i}} \right)^2 + \left(\dfrac{Y(t_i)-Y_i}{\sigma_{Y_i}} \right)^2 \,\right] \,,
\end{equation}
where $N$ is the number of observations, $t_i$ is the $i$th observation time, $(X_i\pm\sigma_{X_i}, Y_i\pm\sigma_{Y_i})\,$ is the $i$th observed hot spot location with its associated error bars, $(X(t_i), Y(t_i))$ is the calculated flux centroid position, and $N_f$ is the number of free parameters for each model. 
In particular, all models have six parameters in common: two for the orientation of the observer $(\theta_{obs}, \phi_{obs})$, two for the black hole's location in the sky $(x_{{\tiny BH}},\, y_{{\tiny BH}})$, and two for the hot spot's initial position on the orbital plane $(x_{spot},\,y_{spot})$. 
We searched for values of $\chi_r^2$ close to unity\footnote{In the process of finding the best-fit value for $\chi_r^2$, we followed a procedure for determining the position of the black hole in the sky.} that indicate a good correlation between the model and observations.
\section{Circular orbit}\label{Sect:Circular}
First and foremost, we considered a circular hot spot trajectory, motivated by the initial fit of \citealt{GRAVITY_2018}. 
The simplistic circular model is further motivated by state-of-the-art numerical simulations of energetic flux tubes in magnetically arrested disks, whose orbits tend to circularize above a radius of $\sim5\,r_g$ from the supermassive black hole (\citealt{Porth2021}).
In the first part of this section, we maintain a constant observer 
angle and black hole spin, whereas in the following subsections, we vary these parameters to assess their impact on the results. 
More specifically, we alter the spin of the black hole in subsection \ref{Subsect:Spi}, and the observation angle and orbital plane inclination in subsection \ref{Subsect:Incline}. \par 
\begin{table}[h]
\caption{Orbital parameters yielding permissible hot spot trajectories for the circular model.}
\begin{center}
 \begin{tabular}{c c c c} 
 \hline \hline 
 Orb. Frequency & Radius & Best-fit  & $\chi_r^2=\chi^2/13$ \\
 $\omega/\omega_K$ &  $r/r_g$ &  $r/r_g$ & $ $ \\
 \hline %\hline
 0.8\tablefootnote{\,We include the best-fit orbital parameters for the sub-Keplerian orbit illustrated in the first panel of Figure \ref{Fig:Circular}, even though it is classified as impermissible.} & 5.5 - 5.8 & 5.8 & 2.015 \\
 1 & 6.5 - 7 & 6.8 & 1.725 \\
 1.2 & 7.3 - 8 & 7.5 & 1.408 \\
 1.5 & 8.5 - 9.5 & 8.8 & 1.02 \\
 1.8 & 9.8 - 10.5 & 10.0 & 1.001 \\
 2 & 10.5 - 11.5 & 11.0 & 1.007 \\ 
 2.2 & 11.5 - 12.2 & 11.8 & 1.01 \\
 2.5 & 12.5 - 13.3 & 13.0 & 1.173 \\
 \hline
\end{tabular}
\end{center}
\tablefoot{\textit{First column}: Orbital frequency of the hot spot. \textit{Second column}: Range of circular radii. \textit{Third column}: Best-fit orbital radius. \textit{Fourth column}: Best-fit $\chi_r^2$ value.}
\label{Table:Circular}
\end{table}
\begin{figure*}[h]
\sidecaption
\includegraphics[width=5.75cm]{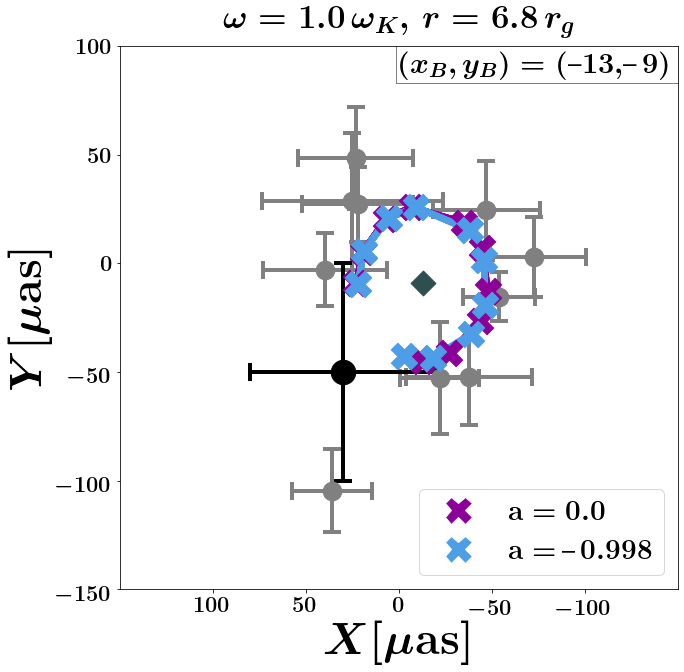} \hspace{0.5cm}
\includegraphics[width=5.35cm]{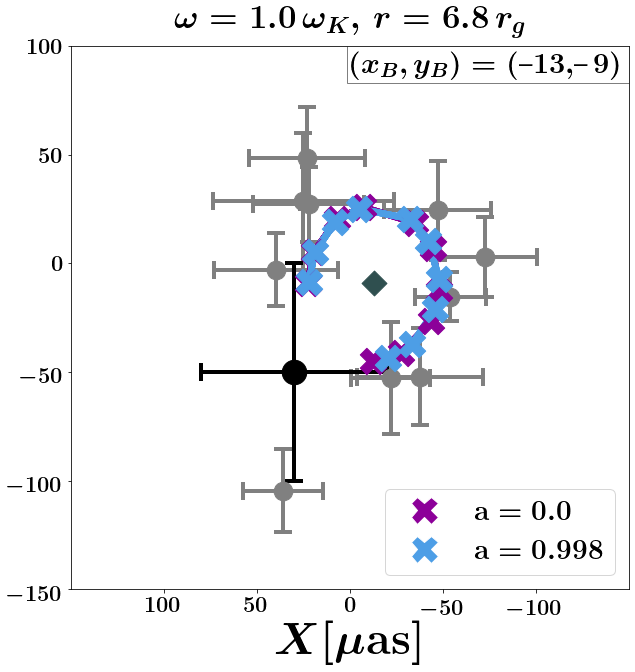} %[width=0.3\linewidth]{GRAVITY_Plots/Spin_Neg.png}
\caption{Circular Keplerian hot spot trajectory illustrated in Figure \ref{Fig:Circular} for a Schwarzschild (purple) and a maximally spinning Kerr black hole (blue) overlapped with the observed flare of July 22, 2018. These orbits share the same black hole position, depicted by a gray diamond and denoted in the top right legend. The bottom right legend indicates the spin of the black hole for each model. The observation angle is face-on and the hot spot rotates \textit{Left Panel}: In the opposite sense as the Kerr black hole ($a=0.998$), \textit{Right Panel}: In the same sense as the Kerr black hole ($a=-0.998$).}
\label{Fig:Spin}
\end{figure*}
In the circular model, we chose a face-on observation angle and a nonrotating black hole. 
The entire orbit lies on the black hole's equatorial plane, as 
is shown below:
\begin{align}
\label{Model:circular_orbit}
x\,(t) &= r_0\,\sin{(\omega t + \phi_0)} \,, \nonumber\\
y\,(t) &= r_0\,\cos{(\omega t + \phi_0)} \,, \\
z\,(t) &= 0 \,, \nonumber
\end{align}  
where $r_0$ and $\phi_0$ are the initial radius and azimuthal angle of the orbit, respectively, and $\omega$ is the orbital frequency of the hot spot. 
There are seven free parameters for this model: the six common to all models and the orbital frequency of the hot spot, ranging from sub-Keplerian to super-Keplerian values.\par
We explored a range of circular radii, spanning from $4\,{\rm r_g}$ to $15\,{\rm r_g}$, and determined the permissible\footnote{\label{Permissible} We classify as permissible orbits, those with no more than two points lying outside the observational error bars.} boundaries for the orbital radius across various orbital frequencies.
Table \ref{Table:Circular} summarizes the parameters that yield permissible trajectories for each orbital frequency, along with the best-fit radius and the associated value of $\chi_r^2$. 
Larger and smaller radii than those reported in the table result in more than two points lying outside the observed error bars and are therefore classified as impermissible. 
Figure \ref{Fig:Circular} illustrates three characteristic hot spot trajectories, corresponding to sub-Keplerian ($0.8\,\omega_K$), Keplerian, and super-Keplerian ($1.8\,\omega_K$) motion, respectively. \par
We notice that faster orbital frequencies require larger circular radii to fit the GRAVITY observations, in line with our expectations. 
The increase in orbital radius is gradual and reflects the rise in angular velocity, which corresponds to the rate at which the hot spot completes the circular orbit. 
For reference, the angular velocity of a hot spot with a super-Keplerian frequency of $2.5\,\omega_K$ orbiting a Schwarzschild black hole at a radius of $12.5\,r_g$ is equal to $\sim0.71c$.
Furthermore, we observe significantly better agreement between the super-Keplerian models and the observed flares, as is evident in the associated values of reduced $\chi^2$ and the third panel in Figure \ref{Fig:Circular}. 
While none of the circular trajectories account for the first and last observed flux-centroid positions (dark red and light purple, respectively), the super-Keplerian orbits lie much closer to the intermediate points due to their larger radii. 
On the other hand, both the Keplerian and sub-Keplerian orbits lie on the inner edge of the observed flare positions and provide a worse fit. \par
Interestingly, we have not identified permissible\footref{Permissible} hot spot trajectories for orbital frequencies beyond $2.5\,\omega_K$. 
Higher angular velocities do not align with the measured time intervals between individual observations and cannot reproduce the observed flaring behavior of July 22.
Similarly, sub-Keplerian orbits require circular radii smaller than $6\,r_g$ and pass through the inner edge of the observed error bars, as was previously mentioned. 
For instance, the sub-Keplerian orbit depicted in the first panel of Figure \ref{Fig:Circular} leaves more than two observed flaring positions unaccounted for. 
Although this orbit is deemed impermissible, we illustrate it for reference purposes. 
Lastly, we note that these results are valid for nearly face-on observation angles, at which one can study the non-deformed orbits.
\subsection{Black hole spin}\label{Subsect:Spi}
In the following paragraphs, we assess the influence of the black hole's spin on the circular orbit given by equation (\ref{Model:circular_orbit}).
It is well known that as the spin of the black hole increases, so does the distortion of the fabric of space-time around it. 
However, these effects are usually limited to regions in close proximity to the black hole, typically of a few gravitational radii, and fade out as one moves away from its sphere of influence. 
\begin{table}[h] 
\caption{Spin values for the Kerr black hole employed in subsection \ref{Subsect:Spi}.}  
\begin{center}                       
\begin{tabular}{c c}  
\hline\hline  
Black Hole Spin &  $\chi_r^2 = \chi^2/12$\\
$a/M$ & $ $ \\
\hline 
-0.998 & 1.748 \\ 
-0.75 & 1.758 \\ 
-0.5 & 1.771 \\ 
-0.25 & 1.806 \\
0 & 1.869 \\              
0.25 & 1.932 \\  
0.5 & 2.025 \\
0.75 & 2.115 \\
0.998 & 2.211 \\
\hline
\end{tabular}
\end{center}
\tablefoot{\textit{First column}: Black hole spin. \textit{Second column}: Value of $\chi_r^2$.}      
\label{Table:Spin} 
\end{table}
To evaluate the impact of selecting a rapidly spinning Kerr black hole compared to its nonrotating Schwarzschild counterpart, we recalculated the Keplerian orbit illustrated in Figure \ref{Fig:Circular} for the range of spin values depicted in Table \ref{Table:Spin}. 
Specifically, we employed a face-on observation angle, a circular radius of $6.8\,r_g$, and a Keplerian orbital velocity, and maintained a fixed black hole position in the sky.
There are eight free parameters for this model: the six common to all models, the orbital frequency of the hot spot, and the spin of the black hole. \par
Figure \ref{Fig:Spin} shows the same hot spot trajectory for both a Schwarzschild (purple line) and a maximally rotating Kerr black hole (blue line) overlapped with the observations.
We discover that when the hot spot's rotation is aligned (anti-aligned) 
with the spin of the black hole, the orbit traces out a slightly larger (smaller) portion of the circle. 
Individual points lag behind when the black hole's rotation opposes them, while the whole orbit is pushed forward when it rotates in the same sense as the black hole. 
In both cases, the effect of the black hole's spin is more prominent in the second half of the trajectory and the average deviation between individual orbits is on the order of 3\%. \par 
It is evident in both panels of Figure \ref{Fig:Spin} that the spin of the black hole does not alter the topology of the orbit itself, but rather slightly aids or delays the hot spot's motion on the desired trajectory. 
While Table \ref{Table:Spin} demonstrates a gradual increase (decrease) in $\chi_r^2$ for positive (negative) black hole spin, the orbit remains virtually the same for all values under investigation.  
We note that the average deviation between the two extreme cases of maximally spinning Kerr black holes ($a=\pm \,0.998M$), depicted in blue in Figure \ref{Fig:Spin}, amounts to approximately 5.5\%. 
\subsection{Inclined orbit}\label{Subsect:Incline}
In this section, we investigate the kinematics of the circular orbit presented in section \ref{Sect:Circular} for different observation angles. 
Specifically, we employed equation (\ref{Model:circular_orbit}) with a Keplerian orbital frequency, as well as an orbital radius of $6.8\,r_g$, and ran our radiative transfer scheme for various observer angles. 
Additionally, we selected a Schwarzschild black hole with a fixed position in the sky.
There are eight free parameters for this model: the six common to all models, the orbital frequency of the hot spot, and the observation angle. 
\par 
It is readily confirmed that for bigger observer angles, %as the observer's inclination increases, 
the orbit deviates from circularity and becomes increasingly elliptical, and therefore covers less surface area on the image plane. 
As a result, the trajectory is pushed to smaller orbital
radii that are incompatible with the observed flaring behavior. 
While one would expect employing a larger initial radius to
solve this issue, values above the limit presented in Table \ref{Table:Circular} are also unfeasible, because the hot
spot cannot traverse the desired orbit in time to account for the observations. \par
Next, we introduced one more free parameter to our investigation, namely, the inclination of the orbital plane.
Once again, we chose a Keplerian orbital velocity, a circular radius of $6.8\,r_g$, and a nonrotating black hole with a fixed position in the sky.
In addition, we rotated the orbit with respect to the x axis\footnote{The observer is located at the positive end of the y-axis, $\phi_{obs}=90\degree{}$.}, as is shown below:
\begin{align}
    \label{Model:inclined_orbit}
    x\sp{\prime}\,(t) &= x\,(t) \,, \nonumber \\
    y\sp{\prime}\,(t) &= y\,(t)\,\cos{i}\ -z\,(t)\,\sin{i} \,, \\
    z\sp{\prime}\,(t) &= y\,(t)\,\sin{i} + z\,(t)\,\cos{i} \,, \nonumber
\end{align}
where $i$ is the aforementioned inclination. 
In this setup, the shape of the orbit remains virtually the same: however, the hot spot's motion is no longer bound to the equatorial plane of the black hole. \par 
We have identified the best-fit values for the orbital plane's inclination for varying observation angles and present our results in Table \ref{Table:Inclined}. 
We have discovered that the optimal orientation for the plane of motion is face-on, meaning that for an observation angle of $\theta_0$, the best fit is recovered for an orbital plane inclination of $-\theta_0$. 
This result holds true for all observation angles and stems from the fact that face-on inclinations produce larger non-deformed orbits, as was previously discussed. 
The goodness of fit is almost identical for the whole range of observer angles, while individual images differ at a scale of 1\%. 
Therefore, a distant observer cannot effectively distinguish face-on orbits at different observation angles.
\begin{table} 
\caption{Model parameters for the inclined hot spot trajectory discussed in subsection \ref{Subsect:Incline}.}  
\begin{center}                        
\begin{tabular}{c c c}  
\hline\hline                 
Observation Angle & Best-fit Inclination & $\chi_r^2 = \chi^2/11$ \\   
\hline
    0\degree{} & 0\degree{} & 2.492 \\                  
   15\degree{} & -15\degree{} & 2.497 \\      
   30\degree{} & -30\degree{} & 2.508 \\ 
   45\degree{} & -45\degree{} & 2.512 \\ 
   60\degree{} & -60\degree{} & 2.526 \\ 
   75\degree{} & -75\degree{} & 2.553 \\ 
   90\degree{} & -90\degree{} & 2.495 \\ 
   \hline
\end{tabular}
\end{center}
\tablefoot{\textit{First column}: Observation angle. \textit{Second column}: Best-fit orbital plane inclination. \textit{Third column}: Best-fit $\chi_r^2$ value.}             
\label{Table:Inclined} 
\end{table}
\begin{table} 
\caption{Orbital parameters investigated for the cylindrical hot spot model in Section \ref{Sect:Cylind}.}
\begin{center}                       
\begin{tabular}{c c c c}  
\hline\hline 
Orb. Frequency & Max. Height & Ej. Velocity & $\chi_r^2 = \chi^2/12$ \\ 
$\omega/\omega_K$ & $z_{max}/r_g$ & $u_z/c$ & $ $ \\
\hline
   0.8 & -10 & -0.12 & 1.802 \\
   0.8 & 0 & 0 & 2.183\\
   0.8 & 10 & 0.12 & 2.353 \\
\hline  
   1 & -10 & -0.12 & 1.609 \\
   1 & -5 & -0.06 & 1.757 \\
   1 & 0 & 0 & 1.869\\
   1 & 5 & 0.06 & 1.943 \\
   1 & 10 & 0.12 & 2.012 \\
\hline
   1.8 & -10 & -0.12 & 0.9785 \\
   1.8 & 0 & 0 & 1.276\\
   1.8 & 10 & 0.12 &1.135 \\
\hline   
\end{tabular}
\end{center}
\tablefoot{\textit{First column}: Orbital frequency of the hot spot. \textit{Second column}: Maximum height of the orbit. \textit{Third column}: Ejection velocity of the hot spot. \textit{Fourth column}: $\chi_r^2$ value.}       
\label{Table:Cylindrical} 
\end{table}
\section{Cylindrical orbit}\label{Sect:Cylind}
In the following sections, we expand our research to include ejected hot spot trajectories, starting with the simple cylindrical model.
The investigation of cylindrical trajectories is motivated by extreme-resolution three-dimensional GRMHD simulations of flaring events in magnetically arrested disks (\citealt{Ripperda2022}).
When a magnetic flux eruption event occurs, the boundary of the rotating disk close to the equator becomes vertical, and thus produces a cylindrical geometry.
In this model,
we chose a face-on observation angle and a Schwarzschild black hole with a fixed position in the sky.
Moreover, we employed the circular trajectory discussed in section \ref{Sect:Circular} and implemented a constant ejection velocity, $u_z$, parallel to the black hole's spin axis, as is shown below:
\begin{align}
\label{Model:cylindrical_orbit}
x\,(t) &= r_0\sin{(\omega t + \phi_0)} \,, \nonumber\\
y\,(t) &= r_0\cos{(\omega t + \phi_0)} \,, \\
z\,(t) &= u_z\,(t - t_0) \,, \nonumber
\end{align}  
where $t_0$ corresponds to the first observational timestamp. 
In this configuration, the hot spot's initial position lies on the equatorial plane of the black hole and it follows a 
cylindrical orbit, either approaching or moving away from the observer, depending on the sign of $u_z$.
It is worth noting that the ejection velocity is selected to ensure that the hot spot reaches its desired maximum height by the end of the observational period.
\begin{figure}[h]
\centering
\includegraphics[width=0.65\linewidth]{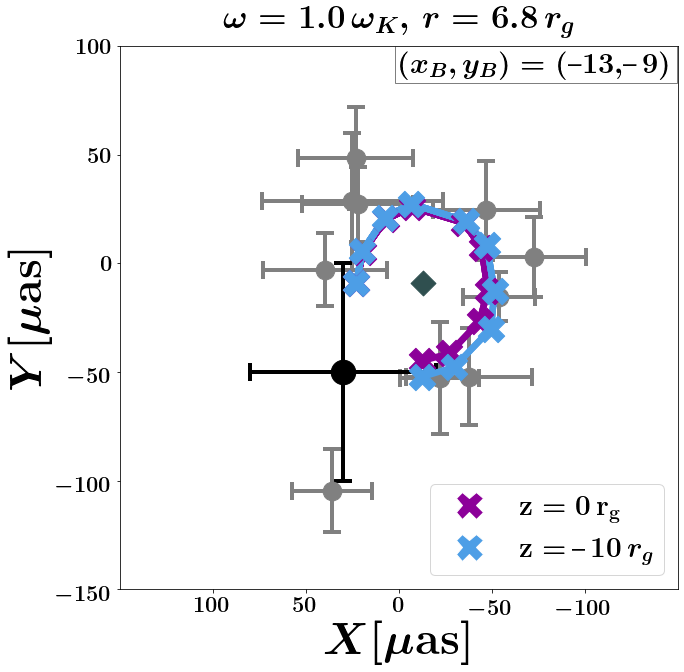} %[width=0.65\linewidth]{GRAVITY_Plots/Cylind_Neg.png}
\caption{Cylindrical hot spot trajectory with a negative ejection velocity (blue) and the circular Keplerian orbit illustrated in Figure \ref{Fig:Circular} (purple) overlapped with the observed flare of July 22, 2018. Both orbits share the same orbital parameters and the bottom right legend indicates the maximum height for each model. The observation angle is face-on and the black hole position is depicted by a gray diamond and denoted in the top right legend.}
 \label{Fig:Cylindrical}
\end{figure}
%
% CONICAL FIGURE
\begin{figure*}[h]
\centering
\includegraphics[width=0.298\linewidth]{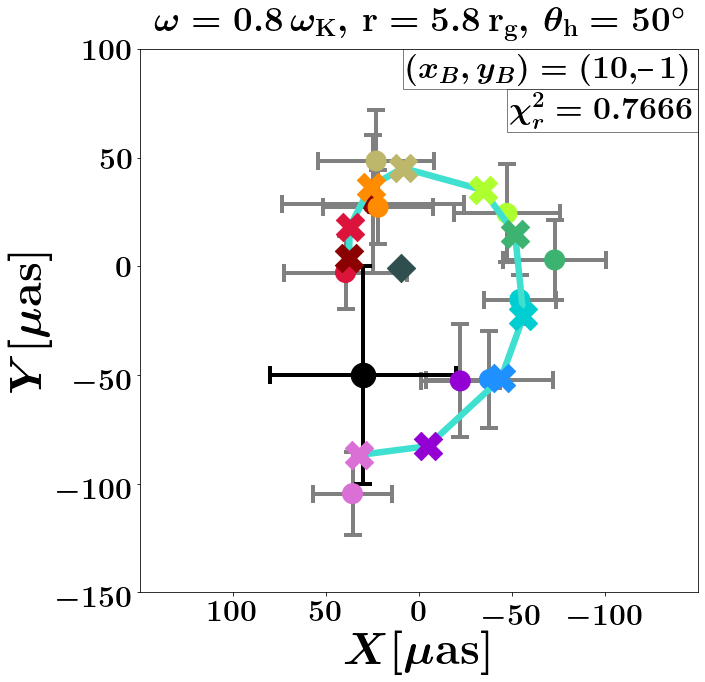} 
\hspace{0.24cm}
\includegraphics[width=0.278\linewidth]{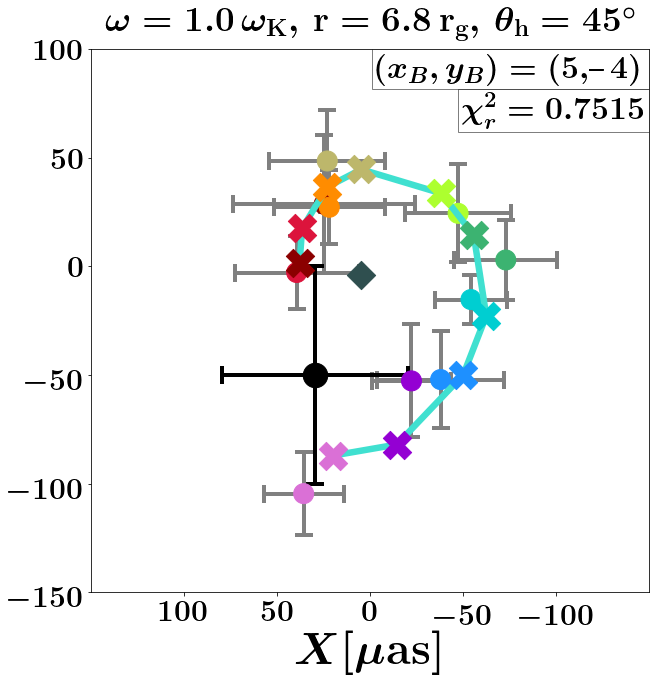}
\hspace{0.24cm}
\includegraphics[width=0.345\linewidth]{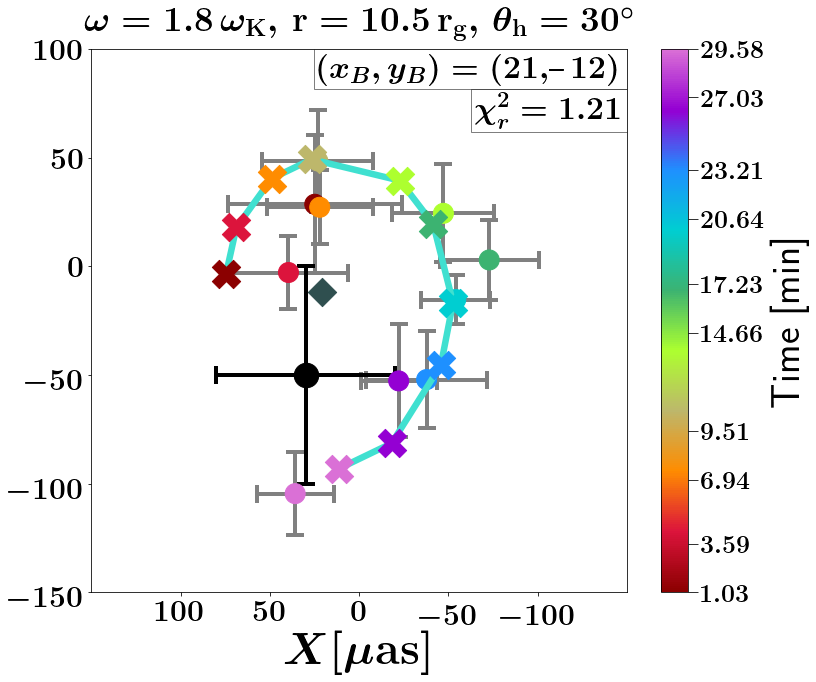}
\caption{Best-fit conical hot spot trajectories for the orbital frequencies depicted in Table \ref{Table:Conical} overlapped with the observed flare of July 22, 2018. Calculated flux-centroid positions (colorful 'x') are color-coordinated with the observations (colorful circles) and the associated timestamps are denoted in the far-right color bar. The approximate position of Sgr$A*$ in the sky is depicted by a black circle. The black hole position for each orbit is illustrated by a dark gray diamond and denoted in the top right legend, along with the associated value of $\chi_r^2$. The observer angle is face-on and the orbital frequency of the hot spot corresponds to \textit{Left Panel}: Sub-Keplerian motion ($0.8\,\omega_K$), \textit{Middle Panel}: Keplerian motion ($\omega_K$),\linebreak and \textit{Right Panel}: Super-Keplerian motion ($1.8\,\omega_K$), respectively.}
 \label{Fig:Conical}
\end{figure*}
Since the trajectory is no longer planar, the cylindrical model possesses an additional degree of freedom, raising the total number to eight: the six common to all models, the orbital frequency, and the ejection velocity of the hot spot.
We present our results in Table \ref{Table:Cylindrical}, organized into three segments according to the orbital frequency of the hot spot. 
The middle row within each segment corresponds to the circular equatorial orbit showcased in Figure \ref{Fig:Circular}, while the rest demonstrate the cylindrical model under the same initial conditions, with heights reaching up to a maximum of $\pm10\,r_g$. \par
We discover that the projection of the cylindrical orbit for positive ejection velocities is nearly identical to its equatorial counterpart, with deviations on the order of $1.5\%$. 
This result holds true for both sub- and super-Keplerian models. 
On the contrary, when a negative ejection velocity is employed, the projection of the trajectory becomes distorted. 
Figure \ref{Fig:Cylindrical} illustrates the comparison between the circular Keplerian orbit depicted in Figure \ref{Fig:Circular} and a cylindrical orbit with identical orbital parameters, featuring an ejection velocity of $-0.12\,c$. 
We notice that the trajectory deviates from circularity and extends to larger radii as the height of the hot spot increases, due to the effect of gravitational lensing.
We observe similar behavior across a broad range of orbital frequencies, although the distortion is less pronounced for distant super-Keplerian models.
\begin{table*}[h]
\caption{Orbital parameters yielding permissible hot spot trajectories for the conical model.}
\begin{center}
 \begin{tabular}{c c c c c} 
 \hline \hline 
 Orb. Frequency & Initial Radius & Opening Angle & Best-fit & $\chi_r^2=\chi^2/10$ \\
 $\omega/\omega_K$ &  $r_0/r_g$ &  $\theta_h$ & $\theta_h$ & $ $ \\
 \hline %\hline
 0.8 & 5.8 & 15\degree\,$-$\,55\degree & 50\degree & 0.7666 \\
 1 & 6.8 & 10\degree\,$-$\,50\degree & 45\degree & 0.7515 \\
 1.8 & 10.5 & 1\degree\,$-$\,35\degree & 30\degree & 1.21 \\
 \hline
\end{tabular}
\end{center}
\tablefoot{\textit{First column}: Orbital frequency of the hot spot. \textit{Second column}: Initial orbital radius. \textit{Third column}: Range of opening angles. \textit{Fourth column}: Best-fit opening angle. \textit{Fifth column}: Best-fit $\chi_r^2$ value.}
\label{Table:Conical}
\end{table*}
As is evident in Table \ref{Table:Cylindrical}, cylindrical models with negative ejection velocities yield a significantly better fit to the data compared to their positive counterparts.\par
In summary, a distant observer cannot effectively differentiate between ejected and planar configurations of the same constant radius, unless the hot spot is moving away from them. 
When the black hole lies between the observer and the emitting hot spot, its gravitational field distorts the path of approaching light rays and alters the resulting image. 
While cylindrical models with positive ejection velocities do not differ from their circular equivalents, those featuring negative velocities result in larger orbital radii and demonstrate a stronger correlation with the observed flaring positions.
However, we harbor doubts regarding whether such an orbit would be observable in practice. 
Given the hot and dense environment around a supermassive black hole like Sgr$A*$, it appears improbable that light from a relatively small aggregation of electrons could penetrate the surrounding disk material.
As a result, in the following sections we limit our analysis to ejected hot spot configurations with exclusively positive ejection velocities.
\section{Conical orbit}\label{Sect:Conical}
State-of-the-art three-dimensional numerical simulations depicting the turbulent environment of accreting black holes demonstrate the formation of hot spots close to the event horizon. 
Magnetic reconnection accelerates the ejected hot spots to relativistic velocities, while the resulting trajectories rotate in a conical manner around the black hole spin axis (\citealt{Nathanail2020,Nathanail2022,ElMellah2023,Lin2024}). 
Motivated by these results, the present
%This  
section explores a conical hot spot model with a fixed opening angle. 
The hot spot maintains a constant orbital frequency and ejection velocity, $u_z$, throughout the trajectory, as is demonstrated below:
\begin{align}
    \label{Model:conical_orbit}
    x\,(t) &= \sin{(\omega t + \phi_0)}\,\tan{\theta_h}\,(z\,(t)+z_0) \,, \nonumber \\
    y\,(t) &= \cos{(\omega t + \phi_0)}\,\tan{\theta_h}\,(z\,(t)+z_0) \,, \\
    z\,(t) &= u_z\,(t + t_{eff}) - z_0 \,, \nonumber
\end{align}
where $\theta_h$ is the opening angle of the cone, and $z_0$ is a constant selected so that the hot spot's initial position lies on the equatorial plane of the black hole.
In this configuration, the effective orbital radius is given by 
\begin{equation}
    \label{reff}
    r_{eff}\,(t) \equiv \tan{\theta_h}\,(z\,(t) + z_0)= \tan{\theta_h}\,u_z\,(t + t_{eff}) \,,
\end{equation}
where $t_{eff}$ is self-consistently determined by the initial orbital radius. \par
In the following paragraphs, we investigate the kinematics of the conical model for various opening angles, employing the orbital frequencies of Figure \ref{Fig:Circular}. 
More specifically, we select a face-on observer angle, a nonrotating Schwarzschild black hole, and an ejection velocity of $0.12\,c$\footnote{The conical trajectories reach a maximum height of $10\,r_g$ by the final observational timestamp.}. 
For all cases, we set the initial radii for the conical orbits close to the best-fit circular radii presented in Section \ref{Sect:Circular}.
Table \ref{Table:Conical} depicts the range of opening angles that result in permissible\footnote{We classify as permissible orbits, those with no more than two points lying outside the observed error bars.} conical trajectories for each orbital frequency. 
It also includes the optimal cone angle along with the associated value of $\chi_r^2$.
We note that the number of free parameters for the conical model is ten: the six common to all models, the orbital frequency and ejection velocity of the hot spot, the opening angle of the cone, and the constant, $z_0$. \par
Orbital frequencies equal to or below the Keplerian value achieve a good correlation with the observed flaring positions for an average cone angle up to $\sim50\degree$. 
On the contrary, super-Keplerian configurations require smaller opening angles, up to $\sim35\degree$. 
It is crucial to emphasize that all investigated hot spot trajectories featuring cone angles at the upper end accurately reproduce the farthest observed flaring position of July 22.  
Conversely, as the opening angle approaches zero, the conical model closely resembles the circular equatorial orbits discussed in Section \ref{Sect:Circular}, with an accuracy of a few percent.
Figure \ref{Fig:Conical} illustrates the best-fit conical orbit for each orbital frequency. 
Remarkably, we notice that frequencies below the Keplerian value exhibit a significantly stronger correlation with the observations of GRAVITY and effectively capture all the observed flux centroid positions. 
On the other hand, super-Keplerian conical orbits sacrifice the fit with intermediate points in order to account for the most distant flaring position.
\par
In summary, the conical model stands out as the first to successfully reproduce the observed flaring behavior across a wide range of orbital frequencies.
While our previous models have demonstrated a good correlation with the observed flares, none have recovered all of the observed flux-centroid positions.
More importantly, the conical model showcases the first sub-Keplerian hot spot trajectory capable of replicating the July 22 flare, with a notable preference toward smaller orbital frequencies.
\section{Parabolic orbit}\label{Sect:Parab}
The final section focuses on parabolic hot spot trajectories, motivated by the parabolic shaped jets demonstrated in three-dimensional numerical simulations of magnetically arrested hot accretion flows (\citealt{Narayan2022}).
While the black hole's spin does not significantly alter the observed hot spot orbit, as is discussed in Section \ref{Subsect:Spi}, it plays a crucial role on the parabolic shape of the jet.
Assuming hot spot formation takes place at the current sheets across the jet boundary (\citealt{Nathanail2014}) and ejected hot spots are then accelerated via magnetic reconnection, different values of black hole spin alter the topology of the hot spot formation sites, and thus significantly affect the ejected hot spot trajectory.
In the parabolic model,
%The final section focuses on parabolic hot spot trajectories around a Schwarzschild black hole. 
%In particular, 
we chose a face-on observation angle and a Schwarzschild black hole with a fixed position in the sky,
and employed a constant orbital frequency and ejection velocity, $u_z$,for the hot spot, as is illustrated below:
\begin{align}
    \label{Model:parabolic_orbit}
    x\,(t) &= \sin{(\omega t + \phi_0)}\,\sqrt{z\,(t)+z_0} \,, \nonumber \\
    y\,(t) &= \cos{(\omega t + \phi_0)}\,\sqrt{z\,(t)+z_0} \,, \\ 
    z\,(t) &= u_z\,(t + t_{eff}) - z_0 \,, \nonumber
\end{align}
where $z_0$ was chosen so that the initial hot spot position rests on the equatorial plane of the black hole and $t_{eff}$ was derived from the initial orbital radius, as is discussed in Section \ref{Sect:Conical}.
In this setup, the effective orbital radius satisfies the parabolic formula \linebreak$z \propto r_{eff}^2$. 
There are nine free parameters for this model: the six common to all models, the orbital frequency and ejection velocity of the hot spot, as well as the constant $z_0$.\par
We searched for permissible\footnote{We classify as permissible orbits, those with no more than two points lying outside the observed error bars.} parabolic trajectories for three distinct orbital frequencies, representing the sub-Keplerian ($0.8\,\omega_K$), Keplerian ($\omega_K$), and super-Keplerian ($1.8\,\omega_K$) regime, respectively. 
The ejection velocity is equal to $0.12\,c$ across all models and the hot spot attains a maximum height of $10\,r_g$ by the end of the observation period.
Table \ref{Table:Parabolic} depicts the range of initial radii yielding permissible hot spot trajectories and the best-fit parameters for each orbital frequency. \par
The parabolic model demonstrates the same range of permissible orbital radii as the circular equatorial orbits investigated in Section \ref{Sect:Circular}\footnote{See the corresponding orbital frequencies in Table \ref{Table:Circular}.}. 
This result stems from the direct relationship between the effective orbital radius and the orbital height of the hot spot, which leads to less pronounced radius variations in the parabolic trajectories.
Furthermore, the average deviation between the parabolic orbits and the circular equatorial trajectories with matching orbital parameters is below $6\%$ in all cases.
Last but not least, super-Keplerian orbital frequencies are in good agreement with the GRAVITY observations, due to their 
\begin{table}[h]
\caption{Orbital parameters yielding permissible hot spot trajectories for the parabolic model.}
\begin{center}
 \begin{tabular}{c c c c c} 
 \hline \hline 
 Orb. Frequency & Initial Radius &  Best-fit & $\chi_r^2=\chi^2/11$ \\
 $\omega/\omega_K$ &  $r_0/r_g$ &  $r_0/r_g$ & $ $ \\
 \hline %\hline
 0.8\tablefootnote{\,As in Section \ref{Sect:Circular}, we include the best-fit orbital parameters for the sub-Keplerian hot spot trajectory for reference.} & 5.5 - 5.8 & 5.8 & 2.239 \\
 1 & 6.5 - 7 & 6.8 & 1.923 \\
 1.8 & 9.8 - 10.5 & 10 & 1.214 \\
 \hline
\end{tabular}
\end{center}
\tablefoot{\textit{First column}: Orbital frequency of the hot spot. \textit{Second column}: Range of initial orbital radii. \textit{Third column}: Best-fit initial radius. \textit{Fourth column}: Best-fit $\chi_r^2$ value.}
\label{Table:Parabolic}
\end{table}
larger orbital radii, whereas sub-Keplerian orbits fail to reproduce the observed flaring behavior of July 22.
\section{Conclusions}\label{Sect:Concl}
Motivated by the intense flaring behavior in our Galactic center and the unprecedented observations of \citealt{GRAVITY_2018}, we employed our recently developed GRRT code to investigate the kinematics of hot spot trajectories in the vicinity of supermassive black holes. 
Our parametric study includes planar and ejected hot spot configurations for both Schwarzschild and Kerr space-times.
In what follows, we outline the key insights gleaned from our research. \newline
(a) Circular hot spot trajectories with super-Keplerian orbital frequency are in line with the radial distance and period of the observed flares and account for all but two of the observed flux-centroid positions.\newline
(b) Circular equatorial orbits featuring sub-Keplerian orbital frequency rest on the inner edge of the observed error bars to account for the desired orbital period and are incompatible with the observations of GRAVITY. \newline
(c) The spin of the black hole does not have a strong impact on the trajectory of the hot spot. Considering maximal spin values either slightly aids or delays hot spot motion by $\sim3\%$. \newline
(d) The observed flares favor face-on observation angles since the hot spot trajectory covers more surface area in the sky plane. In particular, for an observer angle, $\theta_0$, the optimal inclination for the orbital plane is $-\theta_0$. \newline
(e) A distant observer cannot effectively distinguish between Schwarzschild and Kerr black holes or face-on orbits at different observation angles, solely based on the kinematics of the hot spot trajectory. \newline
(f) A cylindrical orbit with a negative ejection velocity extends to larger radii due to the effects of gravitational lensing, whereas an orbit with a positive ejection velocity closely resembles the corresponding circular equatorial trajectory. \newline
(g) The conical model stands out as the first capable of replicating the July 22 flare across a wide range of orbital frequencies. Specifically, conical trajectories featuring sub-Keplerian frequencies exhibit good agreement with the observed flares for opening angles ranging from $10\degree$ to $50\degree$, while super-Keplerian frequencies require opening angles of up to $\sim30\degree$. \newline
(h) Parabolic hot spot trajectories favor super-Keplerian orbital frequencies and closely resemble their circular equatorial equivalents, with average deviations below $6\%$. \newline
\section*{Acknowledgements}
We thank Nicolas Aimar for fruitful discussions regarding general relativistic radiative transfer and our colleagues at the University of Athens for their support throughout this endeavor.
\nocite{*}
\bibliographystyle{aa} 
%\bibliography{bibliography} 
%
\bibliography{parameter_study_for_hot_spot_trajectories_around_SgrA}

\begin{thebibliography}{56}
\expandafter\ifx\csname natexlab\endcsname\relax\def\natexlab#1{#1}\fi

\bibitem[{{Aimar} {et~al.}(2023){Aimar}, {Dmytriiev}, {Vincent}, {El Mellah},
  {Paumard}, {Perrin}, \& {Zech}}]{Aimar2023}
{Aimar}, N., {Dmytriiev}, A., {Vincent}, F.~H., {et~al.} 2023, \aap, 672, A62

\bibitem[{{Baganoff} {et~al.}(2001){Baganoff}, {Bautz}, {Brandt}, {Chartas},
  {Feigelson}, {Garmire}, {Maeda}, {Morris}, {Ricker}, {Townsley}, \&
  {Walter}}]{baganoff2001}
{Baganoff}, F.~K., {Bautz}, M.~W., {Brandt}, W.~N., {et~al.} 2001, \nat, 413,
  45

\bibitem[{Ball {et~al.}(2021)Ball, Özel, Christian, Chan, \&
  Psaltis}]{Ball_2021}
Ball, D., Özel, F., Christian, P., Chan, C.-K., \& Psaltis, D. 2021, The
  Astrophysical Journal, 917, 8

\bibitem[{{Broderick} \& {Loeb}(2005)}]{Broderick2005}
{Broderick}, A.~E. \& {Loeb}, A. 2005, \mnras, 363, 353

\bibitem[{{Chatterjee} {et~al.}(2021){Chatterjee}, {Markoff}, {Neilsen},
  {Younsi}, {Witzel}, {Tchekhovskoy}, {Yoon}, {Ingram}, {van der Klis},
  {Boyce}, {Do}, {Haggard}, \& {Nowak}}]{Chatterjee2021}
{Chatterjee}, K., {Markoff}, S., {Neilsen}, J., {et~al.} 2021, \mnras, 507,
  5281

\bibitem[{{Chen} {et~al.}(2024){Chen}, {Wang}, \& {Yang}}]{Chen2024}
{Chen}, Y., {Wang}, P., \& {Yang}, H. 2024, European Physical Journal C, 84,
  270

\bibitem[{{Dexter} {et~al.}(2020){Dexter}, {Tchekhovskoy},
  {Jim{\'e}nez-Rosales}, {Ressler}, {Baub{\"o}ck}, {Dallilar}, {de Zeeuw},
  {Eisenhauer}, {von Fellenberg}, {Gao}, {Genzel}, {Gillessen}, {Habibi},
  {Ott}, {Stadler}, {Straub}, \& {Widmann}}]{Dexter2020}
{Dexter}, J., {Tchekhovskoy}, A., {Jim{\'e}nez-Rosales}, A., {et~al.} 2020,
  \mnras, 497, 4999

\bibitem[{{Do} {et~al.}(2019){Do}, {Witzel}, {Gautam}, {Chen}, {Ghez},
  {Morris}, {Becklin}, {Ciurlo}, {Hosek}, {Martinez}, {Matthews}, {Sakai}, \&
  {Sch{\"o}del}}]{Do2019}
{Do}, T., {Witzel}, G., {Gautam}, A.~K., {et~al.} 2019, \apjl, 882, L27

\bibitem[{{Dodds-Eden} {et~al.}(2009){Dodds-Eden}, {Porquet}, {Trap},
  {Quataert}, {Haubois}, {Gillessen}, {Grosso}, {Pantin}, {Falcke}, {Rouan},
  {Genzel}, {Hasinger}, {Goldwurm}, {Yusef-Zadeh}, {Clenet}, {Trippe},
  {Lagage}, {Bartko}, {Eisenhauer}, {Ott}, {Paumard}, {Perrin}, {Yuan},
  {Fritz}, \& {Mascetti}}]{dodds-eden2009}
{Dodds-Eden}, K., {Porquet}, D., {Trap}, G., {et~al.} 2009, \apj, 698, 676

\bibitem[{{Eckart} {et~al.}(2006){Eckart}, {Sch{\"o}del}, {Meyer}, {Trippe},
  {Ott}, \& {Genzel}}]{Eckart2006}
{Eckart}, A., {Sch{\"o}del}, R., {Meyer}, L., {et~al.} 2006, \aap, 455, 1

\bibitem[{{El Mellah} {et~al.}(2023){El Mellah}, {Cerutti}, \&
  {Crinquand}}]{ElMellah2023}
{El Mellah}, I., {Cerutti}, B., \& {Crinquand}, B. 2023, \aap, 677, A67

\bibitem[{{Event Horizon Telescope Collaboration} {et~al.}(2022){Event Horizon
  Telescope Collaboration}, {Akiyama}, {Alberdi}, {Alef}, {Algaba}, {Anantua},
  {Asada}, {Azulay}, {Bach}, {Baczko}, {Ball}, {Balokovi{\'c}}, {Barrett},
  {Baub{\"o}ck}, {Benson}, {Bintley}, {Blackburn}, {Blundell}, {Bouman},
  {Bower}, {Boyce}, {Bremer}, {Brinkerink}, {Brissenden}, {Britzen},
  {Broderick}, {Broguiere}, {Bronzwaer}, {Bustamante}, {Byun}, {Carlstrom},
  {Ceccobello}, {Chael}, {Chan}, {Chatterjee}, {Chatterjee}, {Chen}, {Chen},
  {Cheng}, {Cho}, {Christian}, {Conroy}, {Conway}, {Cordes}, {Crawford},
  {Crew}, {Cruz-Osorio}, {Cui}, {Davelaar}, {De Laurentis}, {Deane}, {Dempsey},
  {Desvignes}, {Dexter}, {Dhruv}, {Doeleman}, {Dougal}, {Dzib}, {Eatough},
  {Emami}, {Falcke}, {Farah}, {Fish}, {Fomalont}, {Ford}, {Fraga-Encinas},
  {Freeman}, {Friberg}, {Fromm}, {Fuentes}, {Galison}, {Gammie}, {Garc{\'\i}a},
  {Gentaz}, {Georgiev}, {Goddi}, {Gold}, {G{\'o}mez-Ruiz}, {G{\'o}mez}, {Gu},
  {Gurwell}, {Hada}, {Haggard}, {Haworth}, {Hecht}, {Hesper}, {Heumann}, {Ho},
  {Ho}, {Honma}, {Huang}, {Huang}, {Hughes}, {Ikeda}, {Impellizzeri}, {Inoue},
  {Issaoun}, {James}, {Jannuzi}, {Janssen}, {Jeter}, {Jiang},
  {Jim{\'e}nez-Rosales}, {Johnson}, {Jorstad}, {Joshi}, {Jung}, {Karami},
  {Karuppusamy}, {Kawashima}, {Keating}, {Kettenis}, {Kim}, {Kim}, {Kim},
  {Kim}, {Kino}, {Koay}, {Kocherlakota}, {Kofuji}, {Koch}, {Koyama}, {Kramer},
  {Kramer}, {Krichbaum}, {Kuo}, {La Bella}, {Lauer}, {Lee}, {Lee}, {Leung},
  {Levis}, {Li}, {Lico}, {Lindahl}, {Lindqvist}, {Lisakov}, {Liu}, {Liu},
  {Liuzzo}, {Lo}, {Lobanov}, {Loinard}, {Lonsdale}, {Lu}, {Mao}, {Marchili},
  {Markoff}, {Marrone}, {Marscher}, {Mart{\'\i}-Vidal}, {Matsushita},
  {Matthews}, {Medeiros}, {Menten}, {Michalik}, {Mizuno}, {Mizuno}, {Moran},
  {Moriyama}, {Moscibrodzka}, {M{\"u}ller}, {Mus}, {Musoke}, {Myserlis},
  {Nadolski}, {Nagai}, {Nagar}, {Nakamura}, {Narayan}, {Narayanan},
  {Natarajan}, {Nathanail}, {Fuentes}, {Neilsen}, {Neri}, {Ni}, {Noutsos},
  {Nowak}, {Oh}, {Okino}, {Olivares}, {Ortiz-Le{\'o}n}, {Oyama}, {{\"O}zel},
  {Palumbo}, {Paraschos}, {Park}, {Parsons}, {Patel}, {Pen}, {Pesce},
  {Pi{\'e}tu}, {Plambeck}, {PopStefanija}, {Porth}, {P{\"o}tzl}, {Prather},
  {Preciado-L{\'o}pez}, {Psaltis}, {Pu}, {Ramakrishnan}, {Rao}, {Rawlings},
  {Raymond}, {Rezzolla}, {Ricarte}, {Ripperda}, {Roelofs}, {Rogers}, {Ros},
  {Romero-Ca{\~n}izales}, {Roshanineshat}, {Rottmann}, {Roy}, {Ruiz},
  {Ruszczyk}, {Rygl}, {S{\'a}nchez}, {S{\'a}nchez-Arg{\"u}elles},
  {S{\'a}nchez-Portal}, {Sasada}, {Satapathy}, {Savolainen}, {Schloerb},
  {Schonfeld}, {Schuster}, {Shao}, {Shen}, {Small}, {Sohn}, {SooHoo},
  {Souccar}, {Sun}, {Tazaki}, {Tetarenko}, {Tiede}, {Tilanus}, {Titus},
  {Torne}, {Traianou}, {Trent}, {Trippe}, {Turk}, {van Bemmel}, {van
  Langevelde}, {van Rossum}, {Vos}, {Wagner}, {Ward-Thompson}, {Wardle},
  {Weintroub}, {Wex}, {Wharton}, {Wielgus}, {Wiik}, {Witzel}, {Wondrak},
  {Wong}, {Wu}, {Yamaguchi}, {Yoon}, {Young}, {Young}, {Younsi}, {Yuan},
  {Yuan}, {Zensus}, {Zhang}, {Zhao}, {Zhao}, {Agurto}, {Allardi}, {Amestica},
  {Araneda}, {Arriagada}, {Berghuis}, {Bertarini}, {Berthold}, {Blanchard},
  {Brown}, {C{\'a}rdenas}, {Cantzler}, {Caro}, {Castillo-Dom{\'\i}nguez},
  {Chan}, {Chang}, {Chang}, {Chang}, {Chang}, {Chen}, {Chilson}, {Chuter},
  {Ciechanowicz}, {Colin-Beltran}, {Coulson}, {Crowley}, {Degenaar},
  {Dornbusch}, {Dur{\'a}n}, {Everett}, {Faber}, {Forster}, {Fuchs}, {Gale},
  {Geertsema}, {Gonz{\'a}lez}, {Graham}, {Gueth}, {Halverson}, {Han}, {Han},
  {Hasegawa}, {Hern{\'a}ndez-Rebollar}, {Herrera}, {Herrero-Illana},
  {Heyminck}, {Hirota}, {Hoge}, {Hostler Schimpf}, {Howie}, {Huang}, {Jiang},
  {Jinchi}, {John}, {Kimura}, {Klein}, {Kubo}, {Kuroda}, {Kwon}, {Lacasse},
  {Laing}, {Leitch}, {Li}, {Liu}, {Liu}, {Lin}, {Lu}, {Mac-Auliffe},
  {Martin-Cocher}, {Matulonis}, {Maute}, {Messias}, {Meyer-Zhao},
  {Monta{\~n}a}, {Montenegro-Montes}, {Montgomerie}, {Moreno Nolasco},
  {Muders}, {Nishioka}, {Norton}, {Nystrom}, {Ogawa}, {Olivares}, {Oshiro},
  {P{\'e}rez-Beaupuits}, {Parra}, {Phillips}, {Poirier}, {Pradel}, {Qiu},
  {Raffin}, {Rahlin}, {Ram{\'\i}rez}, {Ressler}, {Reynolds},
  {Rodr{\'\i}guez-Montoya}, {Saez-Madain}, {Santana}, {Shaw}, {Shirkey},
  {Silva}, {Snow}, {Sousa}, {Sridharan}, {Stahm}, {Stark}, {Test},
  {Torstensson}, {Venegas}, {Walther}, {Wei}, {White}, {Wieching}, {Wijnands},
  {Wouterloot}, {Yu}, {Yu (于威)}, \& {Zeballos}}]{EHT2022}
{Event Horizon Telescope Collaboration}, {Akiyama}, K., {Alberdi}, A., {et~al.}
  2022, \apjl, 930, L12

\bibitem[{{Fuerst} \& {Wu}(2004)}]{Fuerst_2004}
{Fuerst}, S.~V. \& {Wu}, K. 2004, \aap, 424, 733

\bibitem[{Genzel(2021)}]{genzel2021}
Genzel, R. 2021, A Forty Year Journey

\bibitem[{{Genzel} {et~al.}(2010){Genzel}, {Eisenhauer}, \&
  {Gillessen}}]{genzel2010}
{Genzel}, R., {Eisenhauer}, F., \& {Gillessen}, S. 2010, Reviews of Modern
  Physics, 82, 3121

\bibitem[{{Genzel} {et~al.}(2003){Genzel}, {Sch{\"o}del}, {Ott}, {Eckart},
  {Alexander}, {Lacombe}, {Rouan}, \& {Aschenbach}}]{genzel2003}
{Genzel}, R., {Sch{\"o}del}, R., {Ott}, T., {et~al.} 2003, \nat, 425, 934

\bibitem[{Ghez {et~al.}(2003)Ghez, Duchêne, Matthews, Hornstein, Tanner,
  Larkin, Morris, Becklin, Salim, Kremenek, Thompson, Soifer, Neugebauer, \&
  McLean}]{Ghez_2003}
Ghez, A.~M., Duchêne, G., Matthews, K., {et~al.} 2003, The Astrophysical
  Journal, 586, L127

\bibitem[{{Gillessen} {et~al.}(2009){Gillessen}, {Eisenhauer}, {Trippe},
  {Alexander}, {Genzel}, {Martins}, \& {Ott}}]{Gillissen2009}
{Gillessen}, S., {Eisenhauer}, F., {Trippe}, S., {et~al.} 2009, \apj, 692, 1075

\bibitem[{{GRAVITY Collaboration} {et~al.}(2017){GRAVITY Collaboration},
  {Abuter}, {Accardo}, {Amorim}, {Anugu}, {{\'A}vila}, {Azouaoui}, {Benisty},
  {Berger}, {Blind}, {Bonnet}, {Bourget}, {Brandner}, {Brast}, {Buron},
  {Burtscher}, {Cassaing}, {Chapron}, {Choquet}, {Cl{\'e}net}, {Collin},
  {Coud{\'e} Du Foresto}, {de Wit}, {de Zeeuw}, {Deen},
  {Delplancke-Str{\"o}bele}, {Dembet}, {Derie}, {Dexter}, {Duvert}, {Ebert},
  {Eckart}, {Eisenhauer}, {Esselborn}, {F{\'e}dou}, {Finger}, {Garcia}, {Garcia
  Dabo}, {Garcia Lopez}, {Gendron}, {Genzel}, {Gillessen}, {Gonte}, {Gordo},
  {Grould}, {Gr{\"o}zinger}, {Guieu}, {Haguenauer}, {Hans}, {Haubois}, {Haug},
  {Haussmann}, {Henning}, {Hippler}, {Horrobin}, {Huber}, {Hubert}, {Hubin},
  {Hummel}, {Jakob}, {Janssen}, {Jochum}, {Jocou}, {Kaufer}, {Kellner},
  {Kendrew}, {Kern}, {Kervella}, {Kiekebusch}, {Klein}, {Kok}, {Kolb}, {Kulas},
  {Lacour}, {Lapeyr{\`e}re}, {Lazareff}, {Le Bouquin}, {L{\`e}na}, {Lenzen},
  {L{\'e}v{\^e}que}, {Lippa}, {Magnard}, {Mehrgan}, {Mellein}, {M{\'e}rand},
  {Moreno-Ventas}, {Moulin}, {M{\"u}ller}, {M{\"u}ller}, {Neumann}, {Oberti},
  {Ott}, {Pallanca}, {Panduro}, {Pasquini}, {Paumard}, {Percheron}, {Perraut},
  {Perrin}, {Pfl{\"u}ger}, {Pfuhl}, {Phan Duc}, {Plewa}, {Popovic}, {Rabien},
  {Ram{\'\i}rez}, {Ramos}, {Rau}, {Riquelme}, {Rohloff}, {Rousset},
  {Sanchez-Bermudez}, {Scheithauer}, {Sch{\"o}ller}, {Schuhler}, {Spyromilio},
  {Straubmeier}, {Sturm}, {Suarez}, {Tristram}, {Ventura}, {Vincent},
  {Waisberg}, {Wank}, {Weber}, {Wieprecht}, {Wiest}, {Wiezorrek}, {Wittkowski},
  {Woillez}, {Wolff}, {Yazici}, {Ziegler}, \& {Zins}}]{GRAVITY2017}
{GRAVITY Collaboration}, {Abuter}, R., {Accardo}, M., {et~al.} 2017, \aap, 602,
  A94

\bibitem[{{GRAVITY Collaboration} {et~al.}(2018{\natexlab{a}}){GRAVITY
  Collaboration}, {Abuter}, {Amorim}, {Anugu}, {Baub{\"o}ck}, {Benisty},
  {Berger}, {Blind}, {Bonnet}, {Brandner}, {Buron}, {Collin}, {Chapron},
  {Cl{\'e}net}, {Coud{\'e} Du Foresto}, {de Zeeuw}, {Deen},
  {Delplancke-Str{\"o}bele}, {Dembet}, {Dexter}, {Duvert}, {Eckart},
  {Eisenhauer}, {Finger}, {F{\"o}rster Schreiber}, {F{\'e}dou}, {Garcia},
  {Garcia Lopez}, {Gao}, {Gendron}, {Genzel}, {Gillessen}, {Gordo}, {Habibi},
  {Haubois}, {Haug}, {Hau{\ss}mann}, {Henning}, {Hippler}, {Horrobin},
  {Hubert}, {Hubin}, {Jimenez Rosales}, {Jochum}, {Jocou}, {Kaufer}, {Kellner},
  {Kendrew}, {Kervella}, {Kok}, {Kulas}, {Lacour}, {Lapeyr{\`e}re}, {Lazareff},
  {Le Bouquin}, {L{\'e}na}, {Lippa}, {Lenzen}, {M{\'e}rand}, {M{\"u}ler},
  {Neumann}, {Ott}, {Palanca}, {Paumard}, {Pasquini}, {Perraut}, {Perrin},
  {Pfuhl}, {Plewa}, {Rabien}, {Ram{\'\i}rez}, {Ramos}, {Rau},
  {Rodr{\'\i}guez-Coira}, {Rohloff}, {Rousset}, {Sanchez-Bermudez},
  {Scheithauer}, {Sch{\"o}ller}, {Schuler}, {Spyromilio}, {Straub},
  {Straubmeier}, {Sturm}, {Tacconi}, {Tristram}, {Vincent}, {von Fellenberg},
  {Wank}, {Waisberg}, {Widmann}, {Wieprecht}, {Wiest}, {Wiezorrek}, {Woillez},
  {Yazici}, {Ziegler}, \& {Zins}}]{GRAVITY_2018}
{GRAVITY Collaboration}, {Abuter}, R., {Amorim}, A., {et~al.}
  2018{\natexlab{a}}, \aap, 615, L15

\bibitem[{{GRAVITY Collaboration} {et~al.}(2019){GRAVITY Collaboration},
  {Abuter}, {Amorim}, {Baub{\"o}ck}, {Berger}, {Bonnet}, {Brandner},
  {Cl{\'e}net}, {Coud{\'e} Du Foresto}, {de Zeeuw}, {Dexter}, {Duvert},
  {Eckart}, {Eisenhauer}, {F{\"o}rster Schreiber}, {Garcia}, {Gao}, {Gendron},
  {Genzel}, {Gerhard}, {Gillessen}, {Habibi}, {Haubois}, {Henning}, {Hippler},
  {Horrobin}, {Jim{\'e}nez-Rosales}, {Jocou}, {Kervella}, {Lacour},
  {Lapeyr{\`e}re}, {Le Bouquin}, {L{\'e}na}, {Ott}, {Paumard}, {Perraut},
  {Perrin}, {Pfuhl}, {Rabien}, {Rodriguez Coira}, {Rousset}, {Scheithauer},
  {Sternberg}, {Straub}, {Straubmeier}, {Sturm}, {Tacconi}, {Vincent}, {von
  Fellenberg}, {Waisberg}, {Widmann}, {Wieprecht}, {Wiezorrek}, {Woillez}, \&
  {Yazici}}]{GRAVITY2019}
{GRAVITY Collaboration}, {Abuter}, R., {Amorim}, A., {et~al.} 2019, \aap, 625,
  L10

\bibitem[{{GRAVITY Collaboration} {et~al.}(2023){GRAVITY Collaboration},
  {Abuter, R.}, {Aimar, N.}, {Bauböck, M.}, {Berger, J. P.}, {Bonnet, H.},
  {Brandner, W.}, {Clénet, Y.}, {Coudé du Foresto, V.}, {de Zeeuw, P. T.},
  {Deen, C.}, {Dexter, J.}, {Duvert, G.}, {Eckart, A.}, {Eisenhauer, F.},
  {Förster Schreiber, N. M.}, {Garcia, P.}, {Gao, F.}, {Gendron, E.}, {Genzel,
  R.}, {Gillessen, S.}, {Guajardo, P.}, {Habibi, M.}, {Haubois, X.}, {Henning,
  Th.}, {Hippler, S.}, {Horrobin, M.}, {Huber, A.}, {Jiménez-Rosales, A.},
  {Jocou, L.}, {Kervella, P.}, {Lacour, S.}, {Lapeyrère, V.}, {Lazareff, B.},
  {Le Bouquin, J.-B.}, {Léna, P.}, {Lippa, M.}, {Ott, T.}, {Panduro, J.},
  {Paumard, T.}, {Perraut, K.}, {Perrin, G.}, {Pfuhl, O.}, {Plewa, P. M.},
  {Rabien, S.}, {Rodríguez-Coira, G.}, {Rousset, G.}, {Sternberg, A.},
  {Straub, O.}, {Straubmeier, C.}, {Sturm, E.}, {Tacconi, L. J.}, {Vincent,
  F.}, {von Fellenberg, S.}, {Waisberg, I.}, {Widmann, F.}, {Wieprecht, E.},
  {Wiezorrek, E.}, {Woillez, J.}, \& {Yazici, S.}}]{GRAVITY2023}
{GRAVITY Collaboration}, {Abuter, R.}, {Aimar, N.}, {et~al.} 2023, A\&A, 677,
  L10

\bibitem[{{GRAVITY Collaboration} {et~al.}(2018{\natexlab{b}}){GRAVITY
  Collaboration}, {Abuter, R.}, {Amorim, A.}, {Bauböck, M.}, {Berger, J. P.},
  {Bonnet, H.}, {Brandner, W.}, {Clénet, Y.}, {Coudé du Foresto, V.}, {de
  Zeeuw, P. T.}, {Deen, C.}, {Dexter, J.}, {Duvert, G.}, {Eckart, A.},
  {Eisenhauer, F.}, {Förster Schreiber, N. M.}, {Garcia, P.}, {Gao, F.},
  {Gendron, E.}, {Genzel, R.}, {Gillessen, S.}, {Guajardo, P.}, {Habibi, M.},
  {Haubois, X.}, {Henning, Th.}, {Hippler, S.}, {Horrobin, M.}, {Huber, A.},
  {Jiménez-Rosales, A.}, {Jocou, L.}, {Kervella, P.}, {Lacour, S.},
  {Lapeyrère, V.}, {Lazareff, B.}, {Le Bouquin, J.-B.}, {Léna, P.}, {Lippa,
  M.}, {Ott, T.}, {Panduro, J.}, {Paumard, T.}, {Perraut, K.}, {Perrin, G.},
  {Pfuhl, O.}, {Plewa, P. M.}, {Rabien, S.}, {Rodríguez-Coira, G.}, {Rousset,
  G.}, {Sternberg, A.}, {Straub, O.}, {Straubmeier, C.}, {Sturm, E.}, {Tacconi,
  L. J.}, {Vincent, F.}, {von Fellenberg, S.}, {Waisberg, I.}, {Widmann, F.},
  {Wieprecht, E.}, {Wiezorrek, E.}, {Woillez, J.}, \& {Yazici,
  S.}}]{GRAVITY2018}
{GRAVITY Collaboration}, {Abuter, R.}, {Amorim, A.}, {et~al.}
  2018{\natexlab{b}}, A\&A, 618, L10

\bibitem[{{GRAVITY Collaboration} {et~al.}(2020{\natexlab{a}}){GRAVITY
  Collaboration}, {Baub{\"o}ck}, {Dexter}, {Abuter}, {Amorim}, {Berger},
  {Bonnet}, {Brandner}, {Cl{\'e}net}, {Coud{\'e} Du Foresto}, {de Zeeuw},
  {Duvert}, {Eckart}, {Eisenhauer}, {F{\"o}rster Schreiber}, {Gao}, {Garcia},
  {Gendron}, {Genzel}, {Gerhard}, {Gillessen}, {Habibi}, {Haubois}, {Henning},
  {Hippler}, {Horrobin}, {Jim{\'e}nez-Rosales}, {Jocou}, {Kervella}, {Lacour},
  {Lapeyr{\`e}re}, {Le Bouquin}, {L{\'e}na}, {Ott}, {Paumard}, {Perraut},
  {Perrin}, {Pfuhl}, {Rabien}, {Rodriguez Coira}, {Rousset}, {Scheithauer},
  {Stadler}, {Sternberg}, {Straub}, {Straubmeier}, {Sturm}, {Tacconi},
  {Vincent}, {von Fellenberg}, {Waisberg}, {Widmann}, {Wieprecht}, {Wiezorrek},
  {Woillez}, \& {Yazici}}]{GRAVITY2020}
{GRAVITY Collaboration}, {Baub{\"o}ck}, M., {Dexter}, J., {et~al.}
  2020{\natexlab{a}}, \aap, 635, A143

\bibitem[{{GRAVITY Collaboration} {et~al.}(2020{\natexlab{b}}){GRAVITY
  Collaboration}, {Jiménez-Rosales, A.}, {Dexter, J.}, {Widmann, F.},
  {Bauböck, M.}, {Abuter, R.}, {Amorim, A.}, {Berger, J. P.}, {Bonnet, H.},
  {Brandner, W.}, {Clénet, Y.}, {de Zeeuw, P. T.}, {Eckart, A.}, {Eisenhauer,
  F.}, {Förster Schreiber, N. M.}, {Garcia, P.}, {Gao, F.}, {Gendron, E.},
  {Genzel, R.}, {Gillessen, S.}, {Habibi, M.}, {Haubois, X.}, {Heißel, G.},
  {Henning, T.}, {Hippler, S.}, {Horrobin, M.}, {Jochum, L.}, {Jocou, L.},
  {Kaufer, A.}, {Kervella, P.}, {Lacour, S.}, {Lapeyrère, V.}, {Le Bouquin,
  J.-B.}, {Léna, P.}, {Nowak, M.}, {Ott, T.}, {Paumard, T.}, {Perraut, K.},
  {Perrin, G.}, {Pfuhl, O.}, {Rodríguez-Coira, G.}, {Shangguan, J.},
  {Scheithauer, S.}, {Stadler, J.}, {Straub, O.}, {Straubmeier, C.}, {Sturm,
  E.}, {Tacconi, L. J.}, {Vincent, F.}, {von Fellenberg, S.}, {Waisberg, I.},
  {Wieprecht, E.}, {Wiezorrek, E.}, {Woillez, J.}, {Yazici, S.}, \& {Zins,
  G.}}]{GRAVITY2020polarimetry}
{GRAVITY Collaboration}, {Jiménez-Rosales, A.}, {Dexter, J.}, {et~al.}
  2020{\natexlab{b}}, A\&A, 643, A56

\bibitem[{{Hornstein} {et~al.}(2007){Hornstein}, {Matthews}, {Ghez}, {Lu},
  {Morris}, {Becklin}, {Rafelski}, \& {Baganoff}}]{hornstein2007}
{Hornstein}, S.~D., {Matthews}, K., {Ghez}, A.~M., {et~al.} 2007, \apj, 667,
  900

\bibitem[{{Huang} {et~al.}(2024){Huang}, {Zhang}, {Guo}, \& {Chen}}]{Huang2024}
{Huang}, J., {Zhang}, Z., {Guo}, M., \& {Chen}, B. 2024, arXiv e-prints,
  arXiv:2402.16293

\bibitem[{{Kocherlakota} {et~al.}(2024){Kocherlakota}, {Rezzolla}, {Roy}, \&
  {Wielgus}}]{kocherlakota2024}
{Kocherlakota}, P., {Rezzolla}, L., {Roy}, R., \& {Wielgus}, M. 2024, arXiv
  e-prints, arXiv:2403.08862

\bibitem[{{Lin} {et~al.}(2023){Lin}, {Li}, \& {Yuan}}]{Lin2023}
{Lin}, X., {Li}, Y.-P., \& {Yuan}, F. 2023, \mnras, 520, 1271

\bibitem[{{Lin} \& {Yuan}(2024)}]{Lin2024}
{Lin}, X. \& {Yuan}, F. 2024, \mnras, 531, 3136

\bibitem[{Marrone {et~al.}(2006)Marrone, Moran, Zhao, \& Rao}]{Marrone_2007}
Marrone, D.~P., Moran, J.~M., Zhao, J.-H., \& Rao, R. 2006, The Astrophysical
  Journal, 654, L57

\bibitem[{Matsumoto {et~al.}(2020)Matsumoto, Chan, \& Piran}]{Matsumoto_2020}
Matsumoto, T., Chan, C.-H., \& Piran, T. 2020, Monthly Notices of the Royal
  Astronomical Society, 497, 2385

\bibitem[{{Narayan} {et~al.}(2022){Narayan}, {Chael}, {Chatterjee}, {Ricarte},
  \& {Curd}}]{Narayan2022}
{Narayan}, R., {Chael}, A., {Chatterjee}, K., {Ricarte}, A., \& {Curd}, B.
  2022, \mnras, 511, 3795

\bibitem[{{Nathanail} \& {Contopoulos}(2014)}]{Nathanail2014}
{Nathanail}, A. \& {Contopoulos}, I. 2014, \apj, 788, 186

\bibitem[{{Nathanail} {et~al.}(2020){Nathanail}, {Fromm}, {Porth}, {Olivares},
  {Younsi}, {Mizuno}, \& {Rezzolla}}]{Nathanail2020}
{Nathanail}, A., {Fromm}, C.~M., {Porth}, O., {et~al.} 2020, \mnras, 495, 1549

\bibitem[{{Nathanail} {et~al.}(2022){Nathanail}, {Mpisketzis}, {Porth},
  {Fromm}, \& {Rezzolla}}]{Nathanail2022}
{Nathanail}, A., {Mpisketzis}, V., {Porth}, O., {Fromm}, C.~M., \& {Rezzolla},
  L. 2022, \mnras, 513, 4267

\bibitem[{Pandya {et~al.}(2016)Pandya, Zhang, Chandra, \& Gammie}]{Pandya_2016}
Pandya, A., Zhang, Z., Chandra, M., \& Gammie, C.~F. 2016, The Astrophysical
  Journal, 822, 34

\bibitem[{Perlick \& Tsupko(2022)}]{Perlick_2022}
Perlick, V. \& Tsupko, O.~Y. 2022, Physics Reports, 947, 1–39

\bibitem[{Ponti {et~al.}(2017)Ponti, George, Scaringi, Zhang, Jin, Dexter,
  Terrier, Clavel, Degenaar, Eisenhauer, Genzel, Gillessen, Goldwurm, Habibi,
  Haggard, Hailey, Harrison, Merloni, Mori, Nandra, Ott, Pfuhl, Plewa, \&
  Waisberg}]{ponti2017}
Ponti, G., George, E., Scaringi, S., {et~al.} 2017, Monthly Notices of the
  Royal Astronomical Society, 468, 2447

\bibitem[{{Porquet} {et~al.}(2003){Porquet}, {Predehl}, {Aschenbach}, {Grosso},
  {Goldwurm}, {Goldoni}, {Warwick}, \& {Decourchelle}}]{porquet2003}
{Porquet}, D., {Predehl}, P., {Aschenbach}, B., {et~al.} 2003, \aap, 407, L17

\bibitem[{{Porth} {et~al.}(2021){Porth}, {Mizuno}, {Younsi}, \&
  {Fromm}}]{Porth2021}
{Porth}, O., {Mizuno}, Y., {Younsi}, Z., \& {Fromm}, C.~M. 2021, \mnras, 502,
  2023

\bibitem[{Pu {et~al.}(2016)Pu, Yun, Younsi, \& Yoon}]{ODYSSEY_Pu_2016}
Pu, H.-Y., Yun, K., Younsi, Z., \& Yoon, S.-J. 2016, The Astrophysical Journal,
  820, 105

\bibitem[{Quataert \& Gruzinov(2000)}]{Quataert_2000}
Quataert, E. \& Gruzinov, A. 2000, The Astrophysical Journal, 545, 842

\bibitem[{{Ripperda} {et~al.}(2022){Ripperda}, {Liska}, {Chatterjee}, {Musoke},
  {Philippov}, {Markoff}, {Tchekhovskoy}, \& {Younsi}}]{Ripperda2022}
{Ripperda}, B., {Liska}, M., {Chatterjee}, K., {et~al.} 2022, \apjl, 924, L32

\bibitem[{Rosa {et~al.}(2022)Rosa, Garcia, Vincent, \& Cardoso}]{Rosa_2022}
Rosa, J.~L., Garcia, P., Vincent, F.~H., \& Cardoso, V. 2022, Physical Review
  D, 106

\bibitem[{{Schnittman} \& {Bertschinger}(2004)}]{Schnittman_2004}
{Schnittman}, J.~D. \& {Bertschinger}, E. 2004, \apj, 606, 1098

\bibitem[{{Sch{\"o}del} {et~al.}(2002){Sch{\"o}del}, {Ott}, {Genzel},
  {Hofmann}, {Lehnert}, {Eckart}, {Mouawad}, {Alexander}, {Reid}, {Lenzen},
  {Hartung}, {Lacombe}, {Rouan}, {Gendron}, {Rousset}, {Lagrange}, {Brandner},
  {Ageorges}, {Lidman}, {Moorwood}, {Spyromilio}, {Hubin}, \&
  {Menten}}]{Schodel2002}
{Sch{\"o}del}, R., {Ott}, T., {Genzel}, R., {et~al.} 2002, \nat, 419, 694

\bibitem[{Shahzadi {et~al.}(2022)Shahzadi, Kološ, Stuchlík, \&
  Habib}]{Shahzadi_2022}
Shahzadi, M., Kološ, M., Stuchlík, Z., \& Habib, Y. 2022, The European
  Physical Journal C, 82

\bibitem[{{Trippe} {et~al.}(2007){Trippe}, {Paumard}, {Ott}, {Gillessen},
  {Eisenhauer}, {Martins}, \& {Genzel}}]{trippe2007}
{Trippe}, S., {Paumard}, T., {Ott}, T., {et~al.} 2007, \mnras, 375, 764

\bibitem[{{Vincent} {et~al.}(2023){Vincent}, {Wielgus}, {Aimar}, {Paumard}, \&
  {Perrin}}]{vincent2023}
{Vincent}, F.~H., {Wielgus}, M., {Aimar}, N., {Paumard}, T., \& {Perrin}, G.
  2023, arXiv e-prints, arXiv:2309.10053

\bibitem[{{Vos} {et~al.}(2022){Vos}, {Mo{\'s}cibrodzka}, \&
  {Wielgus}}]{Vos2022}
{Vos}, J., {Mo{\'s}cibrodzka}, M.~A., \& {Wielgus}, M. 2022, \aap, 668, A185

\bibitem[{{Yfantis} {et~al.}(2023){Yfantis}, {Mo{\'s}cibrodzka}, {Wielgus},
  {Vos}, \& {Jimenez-Rosales}}]{yfantis2023}
{Yfantis}, A.~I., {Mo{\'s}cibrodzka}, M.~A., {Wielgus}, M., {Vos}, J.~T., \&
  {Jimenez-Rosales}, A. 2023, arXiv e-prints, arXiv:2310.07762

\bibitem[{{Younsi}(2014)}]{Younsi_PhD_2014}
{Younsi}, Z. 2014, PhD thesis, University College London, UK

\bibitem[{Younsi \& Wu(2015)}]{Ziri2015}
Younsi, Z. \& Wu, K. 2015, Monthly Notices of the Royal Astronomical Society,
  454, 3283

\bibitem[{{Younsi} {et~al.}(2012){Younsi}, {Wu}, \& {Fuerst}}]{Younsi2012}
{Younsi}, Z., {Wu}, K., \& {Fuerst}, S.~V. 2012, \aap, 545, A13

\bibitem[{Yuan {et~al.}(2003)Yuan, Quataert, \& Narayan}]{Yuan_2003}
Yuan, F., Quataert, E., \& Narayan, R. 2003, The Astrophysical Journal, 598,
  301

\end{thebibliography}

\begin{appendix} 
\section{Code evaluation}\label{Append:Code_Eval}
In the following sections, we assess the validity of our GRRT scheme and present the results of several benchmarking calculations. 
It is essential to clarify that in all calculations performed in this study, the black hole rests at the center of the coordinate system and its mass is normalized to unity.
In addition, the observer is located at a radial distance of $500\,r_g$ from the black hole to alleviate the effects of space-time curvature on the investigated hot spot trajectories.
\subsection{Photon geodesics}\label{Append:Photon_Geod}
\begin{figure}[h]
\centering
\includegraphics[width=0.7\linewidth]{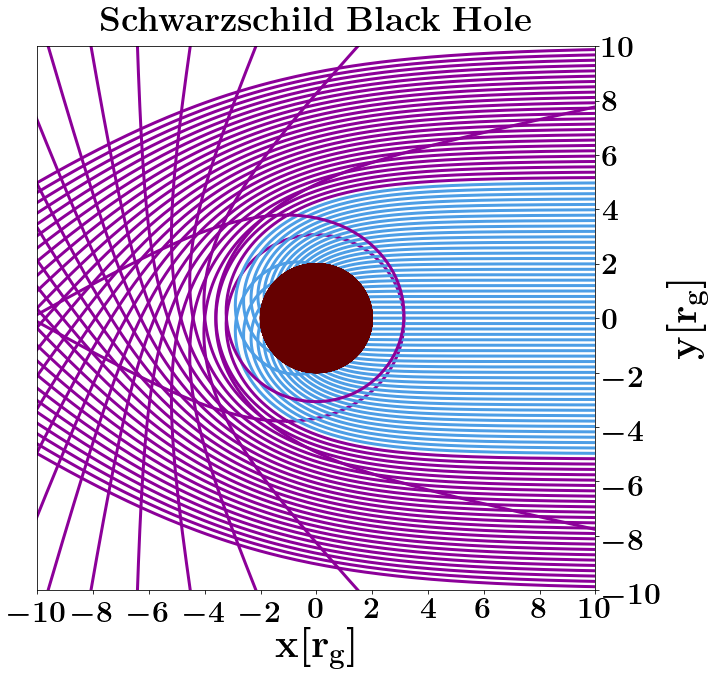} 
\includegraphics[width=0.7\linewidth]{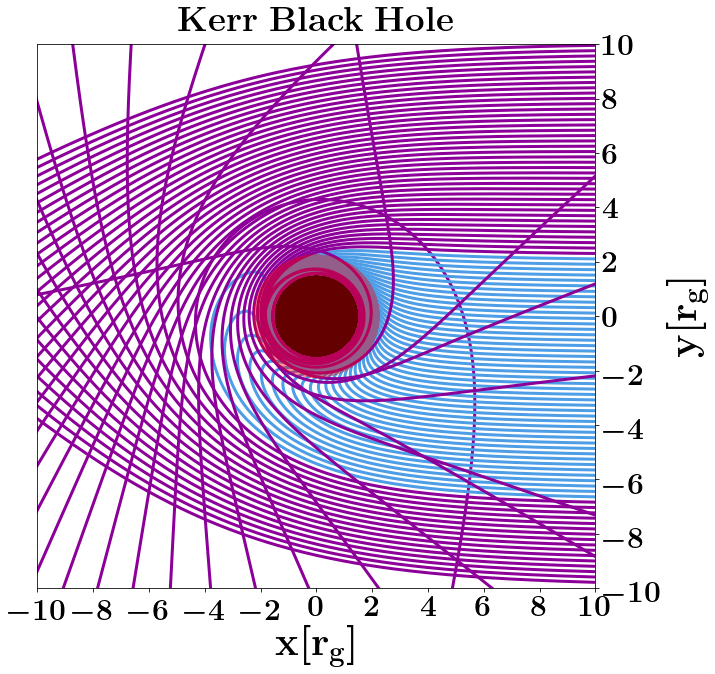} 
\caption{Equatorial photon geodesics around \textit{Top Panel}: a nonrotating Schwarzschild black hole ($a/M=0$), \textit{Bottom Panel}: a maximally rotating Kerr black hole ($a/M=0.998$). The black disk represents the event horizon and the gray ring illustrates the ergosphere of the black hole, respectively. Plunging photon geodesics are depicted in blue, while geodesics that escape to infinity are shown in purple.}
 \label{Fig:Photon_Traj}
\end{figure}

First and foremost, our code tracks the path of light rays within the equatorial plane of a black hole. 
Figure \ref{Fig:Photon_Traj} illustrates 100 photon trajectories in the vicinity of a Schwarzschild (Top Panel) and a maximally rotating Kerr black hole (Bottom Panel). 
The plunging geodesics are shown in blue, whereas the geodesics that escape to infinity are depicted in purple. \par
We readily confirm the spherically symmetric nature of the Schwarzschild space-time, evident from the mirror image relationship between the positive and negative y-axes. 
However, this symmetry breaks down in the Kerr space-time, leading to noticeable alterations in the shadow of the black hole (see Appendix \ref{Append:BH_Shadow}). 
Furthermore, the rapid spin ensures significantly stronger deformation in the photon geodesics compared to the nonrotating case.
\subsection{Black hole shadow}\label{Append:BH_Shadow}
\begin{figure}[h]
\centering
\includegraphics[width=0.7\linewidth]{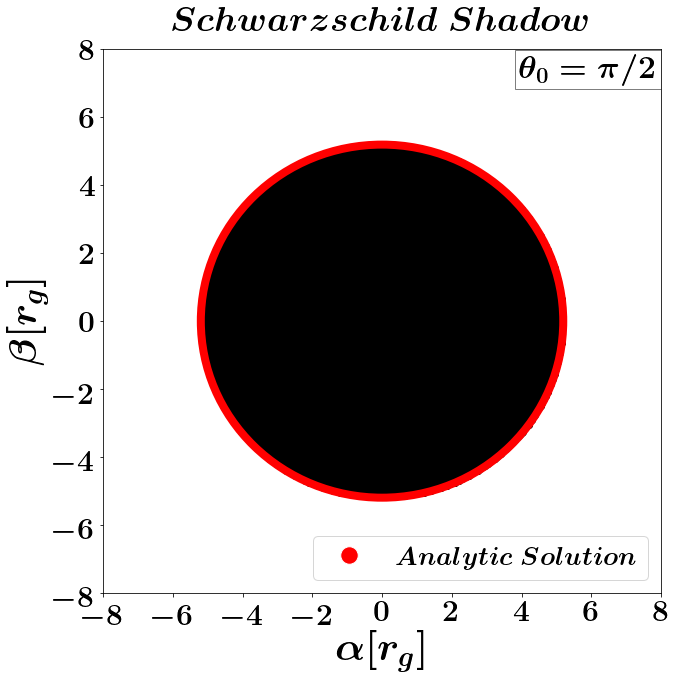} 
\hspace{0.4cm}
\includegraphics[width=0.7\linewidth]{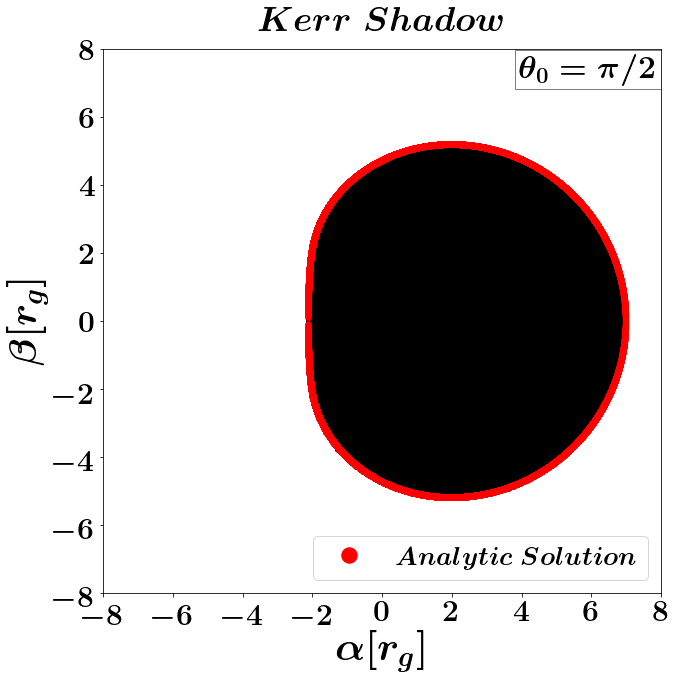} 
\caption{Shadow cast by \textit{Top Panel}: a nonrotating Schwarzschild ($a/M=0$), \textit{Bottom Panel}: a maximally rotating Kerr black hole ($a/M=0.998$), for an observation angle of $90\degree$. The black points represent the computed plunging photon geodesics from our code, while the red line depicts the corresponding analytic solution for each case.}
 \label{Fig:BH_Shadow}
\end{figure}

In this section, we calculate the photon capture region of a black hole. 
Figure \ref{Fig:BH_Shadow} depicts the shadow of a Schwarzschild (Top Panel) and a maximally rotating Kerr black hole (Bottom Panel) %discussed in Appendix A.1 
for a distant observer on the equatorial plane. 
In particular, the black hole shadow is comprised of all the plunging geodesics illustrated by blue lines in the respective panels of Figure \ref{Fig:Photon_Traj}. \par
On the theoretical regime, the shadow cast by a Schwarzschild black hole is spherical, regardless of the observation angle, and given by the following expression (\citealt{ODYSSEY_Pu_2016, Perlick_2022})
\begin{equation}
    \label{Schwarz_Shadow}
    \alpha^2 + \beta^2 = 27\,M^2 \,,
\end{equation}
where $(\alpha, \beta)$ are the celestial coordinates on the image frame of the observer and $M$ is the mass of the black hole.\par
On the other hand, the shadow of a Kerr black hole strongly depends on the observation angle, and its analytic solution for a distant observer is given by (\citealt{Perlick_2022})
\begin{align}
    \label{Kerr_Shadow}
    \alpha &= -\,\frac{\xi(r)}{\sin\theta_0} \,, \\% \:\: , \quad 
    \beta &= \pm\, \sqrt{\eta(r) + a^2\cos^2\theta_0 -\xi ^2(r)\cot^2\theta_0} \,,
\end{align}
where $\theta_0$ is the observer angle and the expressions for $\xi(r)$ and $\eta(r)$ are provided below
\begin{align}
    %\label{Kerr_Shadow_func}
    \xi\,(r) &= \frac{r^2\,(3M-r) -a^2\,(M+r)}{a\,(r-M)} \,, \\
    \eta\,(r) &= \frac{r^3\left(4a^2M - r\,(3M-r)^2\right)}{a^2\,(r-M)^2} \,.
\end{align}
The theoretical solutions (\ref{Schwarz_Shadow}) and (\ref{Kerr_Shadow}) are depicted by red lines in the corresponding panels of Figure \ref{Fig:BH_Shadow}. 
We find excellent agreement between the results of our code and the analytic expressions, underscoring the precision of our calculations.
\subsection{Keplerian hot spot}\label{Append:Kepler_Spot}
In the next section, we consider a spherical hot spot of radius $0.5\,r_g$ rotating in a clockwise direction on the equatorial plane of a Schwarzschild black hole. 
The center of the hot spot is located at the innermost stable circular orbit, corresponding to $6\,r_g$ for a nonrotating black hole, and its four-momentum is given by the following equations (\citealt{Schnittman_2004})
\begin{align}
    p_0 &= -\,\frac{r^2-2Mr\pm a\sqrt{Mr}}{r\left(r^2-3Mr\pm 2a\sqrt{Mr}\right)^{1/2}\,} \,, \label{p0_spot} \\[2pt]
    p_{\phi} &= \pm\,\frac{\sqrt{Mr}\left(r^2\mp 2a\sqrt{Mr} + a^2\right)}{r\left(r^2-3Mr\pm 2a\sqrt{Mr}\right)^{1/2}\,} \,, \label{pφ_spot}
\end{align}
where $M$ is the mass of the black hole, $a$ is the black hole's spin and $r$ is the orbital radius of the hot spot.
The expressions above describe a massive particle in a steady-state disk with an orbital frequency equal to
\begin{equation}
    \label{Ω_spot}
    \Omega_{\phi} \equiv \dfrac{p^{\phi}}{p^0} = \dfrac{g^{\mu\phi}p_{\phi}}{g^{\mu 0}p_0} =\: \dfrac{\pm \sqrt{M}}{\,r^{3/2} \pm a\sqrt{M\,}} \,,
\end{equation}
where the plus and minus signs correspond to prograde and retrograde orbits, respectively. 
As a result, the majority of the disk rotates with an approximately Keplerian angular velocity, and the hot spot maintains its spherical shape throughout the rotation. 
Furthermore, the hot spot is considered optically thick, therefore all computations are terminated on the hot spot surface. \par
In this model, the emissivity is a function of the distance from the hot spot center, within a radiative sphere of $4\,r_{spot}\,$, and the positive part of the z-axis, as is shown below (\citealt{Schnittman_2004}): 
\begin{equation}
    \label{Schnitt_Emission}
    j(\vec{x}) \propto \exp{\left(-\dfrac{d^2}{\,2r_{spot}^2\,} \right)}\, , \:\:\, d=\abs{\vec{x}-\vec{x}_{spot}(t)} \,,
\end{equation}
where $\vec{x}_{spot}$ is the position vector for the hot spot, $d$ is the distance from the hot spot center, and $r_{spot}$ is the hot spot radius. \par
\begin{figure}[h]
    \centering
    \includegraphics[width=0.9\linewidth]{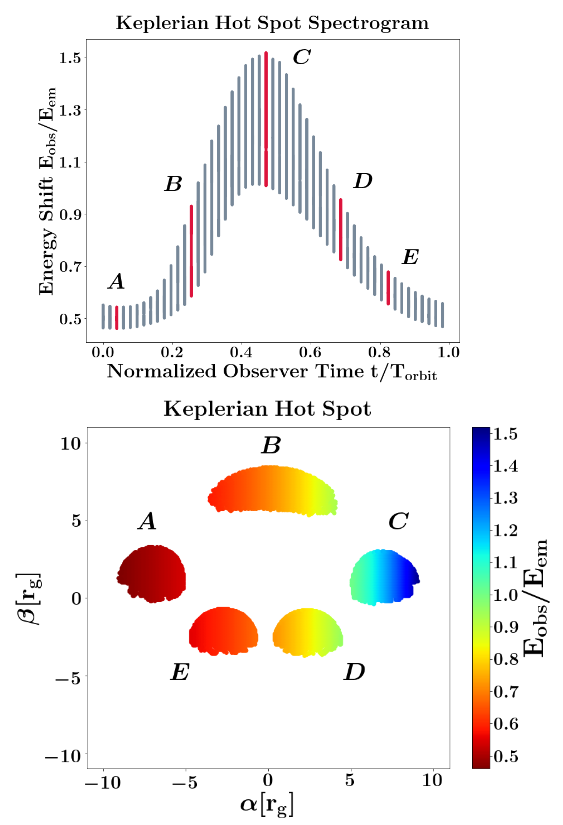}
    \caption{\textit{Top Panel}: Spectrogram of a Keplerian hot spot in a clockwise motion around a Schwarzschild black hole for an observer angle of $60\degree$, \textit{Bottom Panel}: The shape and energy shift of the hot spot at the highlighted timestamps from the Top Panel.}
    \label{fig:Schnitt60}
\end{figure}

 
The Top Panel of Figure \ref{fig:Schnitt60} depicts the spectrogram of the hot spot, namely the evolution of the photons' relative energy shift during rotation to the normalized observer time, for an observation angle of $60\degree$. 
In the Bottom Panel, the shape of the hot spot is illustrated for the five %observer 
timestamps highlighted in red in the spectrogram curve. 
The hot spot begins its motion on the left-hand side of the black hole and the point of maximum red-shift (A), then moves to the far side of the black hole (B), and reaches the point of maximum blue-shift on the right-hand side of the black hole (C). 
Notice the effect of gravitational lensing when the hot spot is directly behind the black hole, resulting in great deformation in its spherical shape.
In summary, our code recovers the expected spectrogram curve and reproduces the shape of the hot spot throughout its rotation (\citealt{Schnittman_2004, ODYSSEY_Pu_2016}). 
\end{appendix}
\end{document}